\documentclass[aps,prd,reprint,amsmath,amssymb,superscriptaddress,floatfix]{revtex4}
\usepackage{graphicx}
\usepackage{amsfonts}
\usepackage{amssymb}
\usepackage{rotating}
\usepackage{booktabs}
\usepackage{xcolor}
\usepackage{soul}
\usepackage{color}
\usepackage{slashed}
\usepackage{multirow}
\usepackage{makecell}
\usepackage{epsf}
\usepackage{ulem}
\usepackage{cancel}
\usepackage{color,bm}
\usepackage[colorinlistoftodos]{todonotes}

\usepackage[colorlinks=true,citecolor=cyan,urlcolor=blue,bookmarks=true,bookmarks=true,bookmarksopen=true,bookmarksnumbered=true,bookmarksopenlevel=3]{hyperref}

\definecolor{airforceblue}{rgb}{0.36, 0.54, 0.66}
\definecolor{steelblue}{rgb}{0.27, 0.51, 0.71}
\definecolor{amber}{rgb}{1.0, 0.49, 0.0}

\def\comment#1{}

\begin{document}

\title{Probing the Sivers asymmetries through $\rm J/\psi$ photoproduction in $\rm p^\uparrow p$ collision with forward proton tagging}
\date{\today}
\author{\textsc{Hao Sun}}
\email{ haosun@mail.ustc.edu.cn\ \ haosun@dlut.edu.cn }
\author{\textsc{Tichouk}}
\author{\textsc{Xuan Luo}}
\affiliation{Institute of Theoretical Physics, School of Physics, Dalian University of Technology, \\ No.2 Linggong Road, Dalian, Liaoning, 116024, P.R.China }

\begin{abstract}
In this paper we probe the Sivers asymmetries through $\rm J/\psi$ photoproduction in $\rm p^\uparrow p$ collision within the non-relativistic QCD framework, based on color octet model and the Transverse Momentum Dependent Parton Distributions (TMDs). Both the DGLAP evolution and the TMD evolution are included. The intensity and the sign of the Sivers asymmetry is strongly related on evolution model used to investigate the Gluon Sivers Function (GSF). A sizable asymmetry is obtained as a function of the rapidity, $\rm log{(x_{\gamma})}$ or $\rm log{(x_{g})}$ dependence using a recent parametrization of GSF at the RHIC and AFTER@LHC experiments with the LHC planned forward detector acceptances.
\end{abstract}
\maketitle
\setcounter{footnote}{0}

\section{INTRODUCTION}
\label{introduction}

The transverse spin physics can be studied in high energy processes that involves the polarized hadrons. This allows us to investigate the polarized quark and gluon structure of the hadrons and provide information on the three-dimensional structure of the nucleons.
The study of transverse spin physics can give more details to QCD dynamics at a high energy scale and is, therefore, of strong interest and highly motivating.

The transverse single-spin asymmetries (SSAs) is one of the topics in spin physics
that drew a lot of attentions since some long time ago \cite{Adams:1991rw,Adams:1991cs}.
The SSA appears in scattering processes when one of the colliding proton is transversely polarized and scatters off an unpolarized proton or a hadron target with respect to the scattering plane. A possible explanation for the presence of the SSA was proposed many years ago which is known as the Sivers effect \cite{Sivers:1989cc}. It considers the nonperturbative quantum correlation between the transverse momentum of partons and the polarization vector of the nucleon, which can be described within the framework of generalized parton model (GPM) \cite{Ji:2004xq,Ji:2004wu}. In the GPM, the inclusive cross section can be written as a convolution of the QCD partonic cross sections, the Transverse Momentum Dependent Partonic Distribution Functions (TMD-PDFs) and the Transverse Momentum Dependent Fragmentation Functions (TMD-FFs) wherein the PDFs and FFs rely on intrinsic momentum $\rm k_{\bot}$ as well as the momentum fraction variable $\rm x$. For more details, see references in \cite{Ji:2004xq,Ji:2004wu,GarciaEchevarria:2011rb,Bacchetta:2006tn,Anselmino:2002pd,Boer:1999mm,Arnold:2008kf,Boer:1997mf,Anselmino:2007fs} in the theoretical aspects toward the understanding of the origin of the SSAs. In the experimental aspects, there has been significant progress in the measurement of the Sivers effects, where they are observed in the experiments by HERMES \cite{Airapetian:2004tw,Airapetian:2009ae,Airapetian:2013bim}, COMPASS \cite{Adolph:2012sp,Adolph:2014fjw,Adolph:2016dvl,Adolph:2017pgv,Aghasyan:2017jop}, JLAB \cite{Qian:2011py,Zhao:2014qvx} and RHIC \cite{Adamczyk:2015gyk} collaborations. The experimental data released by the collaborations have allowed the extraction of the Sivers functions for u and d quarks \cite{Anselmino:2005ea,Anselmino:2016fhz,Anselmino:2016uie,Martin:2017yms}. The gluon sivers function (GSF) has been extracted from SIDIS processes, but still remains poorly measured. An indirect estimation of the GSF exists, which was obtained, within the GPM framework in \cite{DAlesio:2015fwo}, by fitting the midrapidity data on SSA in $\pi^0$ production at RHIC.

Quarkonium production process is an useful tool that is used to probe gluons inside hadron \cite{Brambilla:2010cs} through single photoproductions of $\rm J/\psi$. More recently and more importantly, the study of $\rm J/\psi$ formation has been theoretically carried out in electron-proton ($\rm ep$) \cite{Godbole:2012bx,Mukherjee:2016qxa,Boer:2016bfj,Anselmino:2016fhz,Boer:2016fqd}
and proton-proton ($\rm pp$) \cite{Anselmino:2004nk,DAlesio:2017rzj} collisions. The GSF and linearly polarized gluon distribution \cite{Mukherjee:2015smo,Mukherjee:2016cjw} are studied at length. The mechanism of quarkonium creation out of the two heavy quarks is a nonperturbative process and is treated in terms of different models where the non-relativistic QCD (NRQCD) \cite{Bodwin:1994jh} factorization is one of them that has been chosen because it has effectively explained the $\rm J/\psi$ photoproduction at Tevatron \cite{Abe:1997jz,Acosta:2004yw}, along with data from $\rm J/\psi$ photoproducton at HERA \cite{Adloff:2002ex,Aaron:2010gz,Chekanov:2002at,Abramowicz:2012dh}.
In the NRQCD, the production and decay of heavy quarkonium are split into two steps. To start with, a heavy quark-antiquark pair is perturbatively built at short distances, which is puzzled out by expansion in the strong coupling constant $\rm \alpha_s$. Then, the pair nonperturbatively evolves into quarkonium at a long distance. The short distance coefficients are calculated perturbatively by the projection technique and the long distance matrix elements (LDMEs) are extracted from the experimental data. The LDMEs scale is expanded in powers of v, v being the typical heavy-quark (or antiquark) velocity in the quarkonium rest frame \cite{Lepage:1992tx}. Therefore, the NRQCD factorization can be thought of as doubly expanded expression in terms of $\rm v$ as well as $\rm \alpha_s$. As a matter of fact, the asymmetry is very receptive to the production mechanism. On the one side, in pp collision through $\rm \gamma g$ sub-collision, the final state interactions with the heavy quark and antiquark neutralize among themselves when the pair is produced in a color-singlet configuration, giving a zero asymmetry. On the other side, one gets nonzero asymmetry when the pair is produced in a color-octet configuration \cite{Yuan:2008vn}.

To follow up on the understanding of the Sivers effect origin, numerous launched studies have theoretically been achieved in different key processes in the vein of $\rm ep^\uparrow$ collision, the heavy quark pair and dijet production \cite{Boer:2016fqd}, the inelastic $\rm J/\psi$ photoproduction \cite{Rajesh:2018qks}, $\rm e+p^\uparrow \to e+J/\psi+X$ \cite{Godbole:2012bx,Godbole:2013bca,Godbole:2014tha}, and also in $\rm pp^\uparrow$ scattering for instance $\rm pp^\uparrow \to h+X$\cite{Boer:2015vso}, D-meson production \cite{Anselmino:2004nk,Godbole:2016tvq} and back to back jet correlations \cite{Boer:2003tx}, etc. Even so, a great deal of insufficiencies have been pointed out that certain interactions are in failure in quantifying gluon Sivers function attributable to the problem of TMD factorization breaking contributions \cite{Rogers:2010dm}, some features have been somewhat probed. A plan \cite{Albrow:2008pn} for the study of standard model physics with forward detector bringing in the search of new physics signal outputs are suggested by the FP420 R\&D Collaboration in 2009. To reach this new realm of this interest, detectors in the LHC tunnel need to be readjusted so as to precisely measure very forward protons. The forward detector equipment is relevant for the study of photoproduction process which can exclude many serious backgrounds and the potential of forward proton tagging would give a clean situation for new physics domains. Moreover, the proton-proton collision data will offer knowledge about unexplored phase space areas. Three different forward detector acceptances are given as follows $0.1<\xi<0.5$ , $0.0015<\xi<0.5$ and $0.015<\xi<0.15$, and the all range of the forward detector acceptance  without any cut is $0<\xi<1$. Amongst the hadronic collisions, the processes having one single $\rm J/\psi$, one intact unpolarized hadron emitting photon in the final state would, in any cases, be safe \cite{DAlesio:2015fwo} in measuring the GSF with forward detector acceptances. Henceforth, single heavy quarkonium productions are considered to be clean as probes of the GSF.

In this paper, we delve into the possibility of utilizing single charmonium production to obtain evidences on the Sivers function with the forward detector acceptance together with the presentation of predictions for SSA through the process $\rm hp^{\uparrow}\to h\gamma p^{\uparrow}\to h\mathcal{Q}+X$ where h is a unpolarized hadron and in our case, a proton. The asymmetry has been assessed by employing NRQCD framework within color octet model (COM) owing to the vanishing color octet (CO) contribution in aforementioned pp collision. The unpolarized cross of the single $\rm J/\psi$ production has been calculated to estimate the denominator of SSA. The rapidity distribution of SSA has been estimated in Ref \cite{Goncalves:2017fkt} in DGLAP evolution using Color Evaporation Model (CEM) and we have extended this work to TMD evolution using NRQCD approach. The $\rm y^{J/\psi}$, $\rm log{(x_{\gamma})}$, $\rm log{(x_{g})}$ and $\rm p_{T}^{J/\psi}$ distributions have been evaluated at forward detector in our present work and considerable asymmetries have been observed in NRQCD compared to that one in CEM.

We give estimates on asymmetry for forthcoming suggested experiments at AFTER@LHC which is a fixed target experiment with $\rm \sqrt{s}=115GeV$ and for $\rm \sqrt{s}=200,\ 500GeV$ which will be surveyed at the RHIC with LHC planned forward detector acceptances. Two up-to-date extractions \cite{DAlesio:2015fwo,Anselmino:2016uie} were used for the gluon Sivers function from the SSA data in the pp collision at the RHIC. The paper is structured as follows. The single $\rm J/\psi$ photoproduction with forward proton tagging by using NRQCD and the SSA in DGLAP evolution along with TMD evolution are presented in Sec.\ref{framework}. In Sec.\ref{results}, we give both the input parameters and the numerical results. Discussions and summary are in Sec.\ref{summary}.

\section{CALCULATION FRAMEWORK}
\label{framework}

\subsection{$\rm J/\psi$ photoproduction in $\rm p^\uparrow p$ collision with forward proton tagging}

The fundamental concept of creation of strong electromagnetic fields is originated from the charged proton (p) or charged nucleus (A) moving closely to the speed of light(c). On the one hand, the photon arises from the field of one of the two ultrarelativistic and charged hadron (p or A) can collide with one photon of the other hadron (photon-photon process). On the other hand, this photon can also directly interact with the other hadron (photon-hadron process) \cite{Goncalves:2017fkt}. The total cross section of this process can be split in terms of the equivalent flux of photons into the hadron projectile and the photon-photon or photon-target cross section. At this point, the source of photons presumably comes from the unpolarized hadron (p or A), which interact with the transversely polarized protons at high energies, generating a $\rm J/\psi$ and separating off the proton target.

In the case of $\rm pp^{\uparrow}$ collisions, the process of interest can be separated by tagging the unpolarized proton in the final state, which is present when it emits the photon. We will consider the heavy quarkonium production in the NRQCD factorization formalism. We refer to the heavy quarkonium $\rm J/\psi$ as $\rm \mathcal{Q}$. As a consequence, the hadronic cross section for the $\rm hp^{\uparrow}\to h\gamma p^{\uparrow}\to h\mathcal{Q}+X$ process can be expressed as:
\begin{equation}
\rm \sigma (hp^{\uparrow}\to h\gamma p^{\uparrow}\to h\mathcal{Q}+X)=
\int dx_\gamma d^2{\bold{k}_{\bot\gamma}} f_{\gamma/h}(x_\gamma, \bold{k}_{\bot\gamma})
dx_g d^2{\bold{k}_{\bot g}}
f_{g/p^{\uparrow}}(x_g,{\bold{k}}_{\bot g},\mu_{f})
\sum_{n} \hat{\sigma}(\gamma g\rightarrow Q\overline{Q}[n]+X) \rm \langle 0|\mathcal{O}_{1,8}^{J/\psi }[ n ]|0\rangle
\end{equation}
with $\rm \langle 0|\mathcal{O}_{1,8}^{J/\psi }[n]|0\rangle$ are the long-distance matrix elements, which describe the hadronization of the heavy pair into the physical observable quarkonium state $\rm J/\psi$. The $\rm \widehat{\sigma }(\gamma g\rightarrow Q\overline{Q}[n])$ denotes the short-distance cross section for the partonic process $\rm \gamma g\rightarrow Q\overline{Q}[n]$, which is found by operating the covariant projection method. The Fock states $\rm n$ are given as follows: $\rm ^{1}S_{0}^{[8]},^{3}P_{0}^{[8]},^{3}P_{2}^{[8]}$ for $\rm \gamma g\rightarrow Q\overline{Q}[n]$ partonic process. The final state (h) will be characterized by the presence of one rapidity gap and an intact hadron, which we assume to be the unpolarized one. Both aspects can be used in principle to experimentally separate the vector mesons produced by photon-induced interactions.

In our exploratory study here we will suppose that the transverse momentum dependence of the photon distribution can be described by a simple Gaussian form:
\begin{eqnarray}
\rm f_{\gamma/h}(x_\gamma, \bold{k}_{\bot\gamma}) = f_{\gamma/h}(x_\gamma) \frac{1}{\pi \langle k^2_{\bot\gamma} \rangle} e^{-{\bold{k}}^2_{\bot\gamma} / \langle k^2_{\bot\gamma} \rangle }
\end{eqnarray}
where $\rm x_\gamma$ is the energy fraction of hadron carried by the photon with transverse momentum $\bold{k}_{\bot\gamma}$ and can be symbolized by $\rm x_{\gamma} =\frac{E_{\gamma}}{E}$, the ratio between scattered low-$\rm Q^{2}$ photons $\rm E_{\gamma}$ and incoming energy $\rm E$. The $\rm f_{\gamma /h}(\xi)$ represents the effective photon density function which is defined by the Equivalent Photon Approximation(EPA) \cite{Budnev:1974de,Baur:2001jj} in our computation:
\begin{eqnarray}
\rm f_{\gamma /h}(\xi )=\int_{Q_{\min }^{2}}^{Q_{\max }^{2}}\frac{dN_{\gamma}(\xi )}{d\xi dQ^{2}}dQ^{2},
\end{eqnarray}
where $\rm \frac{dN_{\gamma }(\xi )}{d\xi dQ^{2}}$ is the spectrum of quasi-real photon
\begin{eqnarray}
\rm \frac{dN_{\gamma }(\xi )}{d\xi dQ^{2}}=\frac{\alpha }{\pi }\frac{1}{\xi Q^{2}}[(1-\xi )(1-\frac{Q_{\min }^{2}}{Q^{2}})F_{E}+\frac{\xi ^{2}}{2}F_{M}]
\end{eqnarray}
with
\begin{eqnarray}
\rm Q_{\min }^{2}=\frac{m_{p}^{2}\xi ^{2}}{1-\xi }, \ \ \
F_{E}=\frac{4m_{p}^{2}G_{E}^{2}+Q^{2}G_{M}^{2}}{4m_{p}^{2}+Q^{2}}, \ \ \
G_{E}^{2}=\frac{G_{M}^{2}}{\mu _{p}^{2}}=(1+\frac{Q^{2}}{Q_{0}^{2}})^{-4}, \ \ \
F_{M}=G_{M}^{2},
\end{eqnarray}
where $\rm \alpha$ is the fine-structure constant, $\rm \mu _{p}^{2} = 7.78$ is the magnetic moment of the proton, $\rm{Q_{0}^{2}}$=0.71 GeV$^{2}$, $\rm m_p$ is its mass, the range of $\rm Q_{\max }^{2}$ is valued by around 2 GeV$^{2}$ and $\rm \xi$ represents $\rm x_{\gamma}$. $\rm f_{g/p^{\uparrow(\downarrow)}}(x_{g},\bold{k}_{\bot g},\mu_{f})$ stands for the number density of gluon with light-cone momentum fraction $\rm x_g$ and transverse momentum $\rm \bold{k}_{\bot g}=k_{\bot g}(\cos\phi_a, \sin\phi_a)$ inside the transversely polarized proton. The polarization of  proton is upwards or downwards with respect to the production plane, moving along the $\rm \hat{z}$-axis. Considering the partonic process as $\rm \gamma(p_{1})+g(p_{2})\to Q\overline{Q}[n](p_{3})$, the final total cross section for $\rm hp^{\uparrow}\to h\gamma p^{\uparrow}\to h\mathcal{Q}+X$ process can be expressed as
\begin{eqnarray}\label{2to1} \nonumber
\rm \sigma (hp^{\uparrow}\to h\gamma p^{\uparrow}\to h\mathcal{Q}+X) &=& \rm
\int
\frac{\pi}{s^2 x_g x^2_\gamma} \frac{1}{N_{col}N_{pol}} \overline{\sum }\left\vert\mathcal{ A}_{S,L}\right\vert ^{2}
f_{\gamma/h}(x_\gamma, \bold{k}_{\bot\gamma} ) f_{g/p^\uparrow}(x_g,\bold{k}_{\bot g},\mu) )  \\
&& \rm \langle 0|\mathcal{O}_{1,8}^{J/\psi }[n]|0\rangle d^2\bold{p}_{3} dx_\gamma d^2\bold{k}_{\bot \gamma} \ \ \
\end{eqnarray}
with $\rm x_g$ fixed by $\rm x_g=m_{3T}^2/(sx_\gamma$) and $\rm \bold{k}_{\bot \gamma}$ fixed by $\rm \bold{p}_{3T}-\bold{k}_{\bot g}$. Here $\rm x_\gamma$ is integrated in the region $x_{\gamma\min} < x_\gamma < x_{\gamma\max}$ and $x_{\gamma\min}(x_{\gamma\max})$ is the lower(upper) limit of forward detector acceptance. $\rm m_T$ is the transverse mass of the particle which defined as $\rm m_T=\sqrt{m^2+p_T^2}$. The $\rm s$ and $\rm m$ are respectively the square of center-of-mass energy of collider and the mass of particle. Similarly as in photoproduction induced by electron proton collisions, we can define the z parameter $\rm z = P_{h}\cdot P_{3}/P_{h}\cdot q_{\gamma}$
where P, $\rm q_{\gamma}$ are the momenta of the proton and the virtual photon. The data is taken in  elastic regime for the $\rm \gamma g\rightarrow Q\overline{Q}[n]$ partonic process. Unlike, the inelastic regime is commonly considered to be the area where $\rm z$ is below 0.8 or 0.9. The elastic regime is considered to be the area near $\rm z=1$ which is exactly where we have concentrated on for our $\rm J/\psi$ production.

The summation in Eq.(\ref{2to1}) is taken over the spins and colors of initial and final states,
and the bar over the summation denotes averaging over the spins and colors of initial parton.
$\rm N_{col}$ and $\rm N_{pol}$ refer to as the numbers of colors and polarization of states $\rm n$, separately. In the notation of Ref \cite{Petrelli:1997ge}, we have
\begin{equation}
\begin{split}
\rm \mathcal{A}_{Q\overline{Q}}[^{1}S_{0}^{(1/8)}]=&\rm Tr[\mathcal{C}_{(1/8)}\Pi _{0}\mathcal{A}]_{q=0},\\
\rm \mathcal{A}_{Q\overline{Q}}[^{3}S_{1}^{(1/8)}]=&\rm \epsilon _{\alpha }Tr[\mathcal{C}_{(1/8)}\Pi_{1}^{\alpha }\mathcal{A}]_{q=0},\\
\rm \mathcal{A}_{Q\overline{Q}}[^{1}P_{1}^{(1/8)}]=&\rm \epsilon _{\beta }\frac{d}{dq_{\beta }}Tr[\mathcal{C}_{(1/8)}\Pi _{0}\mathcal{A}]_{q=0},\\
\rm \mathcal{A}_{Q\overline{Q}}[^{1}P_{J}^{(1/8)}]=&\rm \epsilon_{\alpha \beta } ^{(J)}\frac{d}{dq_{\beta }}Tr[\mathcal{C}_{(1/8)}\Pi _{1}^{\alpha }\mathcal{A}]_{q=0},
\end{split}
\end{equation}
where $\rm \mathcal{A}$ denotes the QCD amplitude with amputated heavy-quark spinors, the lower index $\rm q$ represents the momentum of the heavy-quark in the $\rm Q\overline{Q}$ rest frame. $\rm \Pi _{0/1}$ are spin projectors onto spin singlet and spin triplet states stated as
\begin{equation}
\begin{split}
\rm \Pi _{0}=&\rm \rm \frac{1}{\sqrt{8m^{3}}}(\frac{\slashed{P}}{2}-\slashed{q}-m)\gamma _{5}(\frac{\slashed{P}}{2}+\slashed{q}+m),\\
\rm \Pi _{1}^{\alpha }=&\rm \rm \frac{1}{\sqrt{8m^{3}}}(\frac{\slashed{P}}{2}-\slashed{q}-m)\gamma ^{\alpha }(\frac{\slashed{P}}{2}+\slashed{q}+m),
\end{split}
\end{equation}
where $\rm P$ is the total momentum of heavy quarkonium, $\rm q$ is the relative momentum between the $\rm Q\overline{Q}$ pair, and $\rm m_{Q}$ is the mass of heavy quark. $\rm \mathcal{C}_{1/8}$ are color factor projectors onto the color-singlet and color-octet states and can be expressed as follows:
\begin{equation}
\begin{split}
\rm C_{1} =&\rm \frac{\delta _{ij}}{\sqrt{N_{c}}}\\
\rm C_{8} =&\rm \sqrt{2}T_{ij}^{c},
\end{split}
\end{equation}
where $\rm N_{c}$ is the number of color, and $\rm T_{ij}^{c}$ is the generator of $\rm SU(N_{c})$.
The summation over the polarization is given as:
\begin{equation}
\begin{split}
\sum_{J_{z}}\varepsilon _{\alpha }\varepsilon _{\alpha ^{^{\prime }}}^{\ast
} =&\Pi _{\alpha \alpha ^{^{\prime }}},\\
\sum_{J_{z}}\varepsilon _{\alpha \beta }^{0}\varepsilon _{\alpha ^{^{\prime
}}\beta ^{^{\prime }}}^{0\ast } =&\frac{1}{3}\Pi _{\alpha \beta }\Pi
_{\alpha ^{^{\prime }}\beta ^{^{\prime }}},\\
\sum_{J_{z}}\varepsilon _{\alpha \beta }^{1}\varepsilon _{\alpha ^{^{\prime
}}\beta ^{^{\prime }}}^{1\ast } =&\frac{1}{2}(\Pi _{\alpha \alpha
^{^{\prime }}}\Pi _{\beta \beta ^{^{\prime }}}-\Pi _{\alpha \beta ^{^{\prime
}}}\Pi _{\alpha ^{^{\prime }}\beta }),\\
\sum_{J_{z}}\varepsilon _{\alpha \beta }^{2}\varepsilon _{\alpha ^{^{\prime
}}\beta ^{^{\prime }}}^{2\ast } =&\frac{1}{2}(\Pi _{\alpha \alpha
^{^{\prime }}}\Pi _{\beta \beta ^{^{\prime }}}+\Pi _{\alpha \beta
^{^{\prime }}}\Pi _{\alpha ^{^{\prime }}\beta })-\frac{1}{3}\Pi _{\alpha
\beta }\Pi _{\alpha ^{^{\prime }}\beta ^{^{\prime }}},
\label{tesfin}
\end{split}
\end{equation}
where $\rm \varepsilon_{\alpha}$ ($\varepsilon_{\alpha \beta }$) represents the polarization vector (tensor) of the $\rm Q\overline{Q}$ states, $\rm \Pi_{\alpha\beta}=-g_{\alpha\beta}+\frac{P_{\alpha}P_{\beta}}{M^{2}}$ and $\rm M$ is the heavy quarkonium mass. The amplitude squares for $\rm 2\to 1$ partonic processes are presented as follows \cite{Cacciari:1996dg}:
\begin{eqnarray}
\rm
&&\rm \overline{\sum \bigskip }\left\vert  \mathcal{ M}\left[ ^{2S+1}L_{J}^{[1,8]}\right]
\right\vert ^{2}=\frac{1}{N_{col}N_{pol}} \overline{\sum }\left\vert\mathcal{ A}_{S,L}\right\vert ^{2}
\end{eqnarray}
where
\begin{eqnarray} \nonumber
\rm
&&\rm \overline{\sum \bigskip }\left\vert  \mathcal{ M}\left[ ^{1}S_{0}^{[8]}\right] \right\vert ^{2}=\frac{(4\pi
)^{2}\alpha\alpha _{s}e_{c}^{2} }{2M}, \\\nonumber
\rm
&&\rm \overline{\sum \bigskip }\left\vert  \mathcal{ M}\left[ ^{3}P_{0}^{[8]}\right] \right\vert ^{2}=\frac{6(4\pi
)^{2}\alpha \alpha _{s}e_{c}^{2}}{M^{3}}, \\
\rm
&&\rm \overline{\sum \bigskip }\left\vert  \mathcal{ M}\left[ ^{3}P_{2}^{[8]}\right] \right\vert ^{2}=\frac{8(4\pi
)^{2}\alpha\alpha _{s}e_{c}^{2} }{5M^{3}}.
\end{eqnarray}
At low $\rm  p_{T}^{J/\psi}$, the heavy quarkonia is dominantly produced at high energy colliders via color-octet channel. Finally, we have
\begin{equation}
\rm \left\vert \overline{\mathcal{M}}\right\vert ^{2}=(4\pi )^{2}e_{c}^{2}\alpha \alpha
_{s}(\frac{1}{2M}\rm \langle 0|\mathcal{O}_{8}^{J/\psi }(^{1}S_{0})|0\rangle +
\frac{6}{M^{3}}\rm \langle 0|\mathcal{O}_{8}^{J/\psi }(^{3}P_{0})|0\rangle  +
\frac{8}{5M^{3}}\rm \langle 0|\mathcal{O}_{8}^{J/\psi }(^{3}P_{2})|0\rangle) .
\end{equation}

\subsection{Sivers asymmetry and parameterization in DGLAP evolution}

The transverse single spin asymmetries (SSAs) for the process $\rm h + p^{\uparrow} \to J/\psi + X$ is defined by
\begin{eqnarray}\label{AN}
\rm A_N = \frac{d\sigma^\uparrow - d\sigma^\downarrow}{d\sigma^\uparrow + d\sigma^\downarrow} = \frac{d\Delta\sigma}{2d\sigma},
\end{eqnarray}
where $\rm d\sigma^{\uparrow(\downarrow)}$ denotes the single-polarized cross section, in which one of the protons in the initial state is polarized along the transverse direction ${\uparrow(\downarrow)}$ with respect to the production plane. One has that the cross section for the $\rm J/\psi$ photoproduction is proportional to the number density of gluons inside a proton with transverse polarization $\bold{S}_{\bot}$ and momentum $\rm \bold{P}$. We choose the frame where the polarized proton is moving along the z axis with momentum $\rm \bold{P}$ and is transversely polarized with $\rm \bold{S}_{\bot}=S_\bot(\cos\phi_s, \sin\phi_s, 0)$. For a general value of the transverse spin $\rm \bold{S}_\bot$, it is parameterized in terms of the gluon Sivers function (GSF) $\rm \Delta^N f_{g/p^\uparrow}$, as follows
\begin{eqnarray}
\rm f_{g/p^\uparrow}(x_g,\bold{k}_{\bot g},\bold{S}_{\bot},\mu) =
 f_{g/p}(x_g,k_{\bot g},\mu) + \frac{1}{2}\Delta^N f_{g/p^\uparrow}(x_g,k_{\bot g},\mu)\ \bold{{\hat{S}}}_{\bot}\cdot(\bold{\hat{P}}\times\bold{\hat{k}}_{\bot g} ).
\end{eqnarray}
where $\rm f_{g/p}(x_g,k_{\bot g},\mu)$ is the unpolarized Transverse Momentum Dependent(TMD) gluon distribution. It is generally assumed that the unpolarized gluon TMDs obey the Gaussian distribution at low. The spectra appear to have a Gaussian shape. The Gaussian parameterization of an unpolarized TMD \cite{Anselmino:2005ea} which is commonly and phenomenologically used is given by
\begin{eqnarray}
\rm f_{g/p}(x_g, \bold{k}_{\bot g},\mu) = f_{g/p}(x_g,\mu)\frac{1}{\pi \langle k^2_\bot \rangle } e^{-k^2_\bot / \langle k^2_\bot \rangle } .
\end{eqnarray}
Here $\rm f_{g/p}(x_g,\mu)$ is the normal collinear PDF, which is measured at the scale $\rm \mu$.
The collinear PDF obeys the Dokshitzer-Gribov-Lipatov-Altarelli-Parisi(DGLAP) scale evolution.
There is no evolution for normalized Gaussian in the transverse momenta $\rm {k}_\bot $.
The transverse momentum of the initial gluon is $\rm \bold{k}_{\bot g}=k_{\bot g}(\cos\phi_g, \sin\phi_g, 0)$, so that $\rm \bold{{\hat{S}}}_{\bot}\cdot(\bold{\hat{P}}\times\bold{\hat{k}}_{\bot g})=\sin(\phi_g-\phi_s)$. For numerical estimation we can take $\rm \phi_s=\pi/2$. In considering with Eq.({\ref{2to1}}), we can then write the numerator and denominator of Eq.(\ref{AN}) as
\begin{eqnarray} \nonumber
\rm
\frac{d\sigma^\uparrow}{d^2 \bold{p}_{3}} -  \frac{d\sigma^\downarrow}{d^2 \bold{p}_{3}}  &=& \rm
\int dx_\gamma d^2{\bold{k}}_{\bot\gamma} f_{\gamma/h}(x_\gamma, \bold{k}_{\bot\gamma})
[ f_{g/p^{\uparrow}}(x_g, \bold{k}_{\bot g}, \mu) - f_{g/p^{\downarrow}}(x_g, \bold{k}_{\bot g}, \mu)]
\frac{\pi}{s^2x_g x^2_\gamma} |\overline{\mathcal{M}}|^2 \sin(\phi_3-\phi_s) \ , \\
\rm
\frac{d\sigma^\uparrow}{d^2 \bold{p}_{3}} +  \frac{d\sigma^\downarrow}{d^2 \bold{p}_{3}}  &=& \rm
2 \int dx_\gamma d^2{\bold{k}}_{\bot\gamma} f_{\gamma/h}(x_\gamma, \bold{k}_{\bot\gamma})
f_{g/p}(x_g, \bold{k}_{\bot g}) \frac{\pi}{s^2 x_g x^2_\gamma} |\overline{\mathcal{M}}|^2
\end{eqnarray}
where $\rm \sin(\phi_3-\phi_s)$ is a weighted factor and $\rm \phi_3$ and $\rm \phi_s$ are the azimuthal angles of the $\rm J/\psi$ and proton spin, respectively.

And we also have
\begin{eqnarray}
\rm
 \Delta^N f_{g/p^\uparrow} (x_g,{\bold{k}_{\bot g},\mu ) } =
[ f_{g/p^{\uparrow}}(x_g, \bold{k}_{\bot g},\mu) - f_{g/p^{\downarrow}}(x_g, \bold{k}_{\bot g},\mu)]
= \Delta^N f_{g/p^\uparrow}(x_g,k_{\bot g},\mu) \ \bold{\hat{S}}_\bot \cdot (\bold{\hat{P}} \times \bold{\hat{k}}_{\bot g}).
\end{eqnarray}
The parameterization of the gluon sivers function can be described by the well-known Gaussian-like format as follows
\begin{eqnarray}
\rm \Delta^N f_{g/p^\uparrow}(x_g,k_{\bot g},\mu) = 2 N_g(x_g) f_{g/p}(x_g,\mu) h(k_{\bot g})
\frac{ e^{-k^2_{\bot g} / \langle k^2_{\bot g} \rangle} }{\pi \langle k^2_{\bot g} \rangle} ,
\end{eqnarray}
 where 
\begin{eqnarray}\label{Ng}
\rm \mathcal{N}_g(x_g) = N_g x_g^\alpha (1-x_g)^\beta \frac{(\alpha+\beta)^{\alpha+\beta}}{\alpha^\alpha \beta^\beta}
\end{eqnarray}
with $\rm |N_g|\leq 1$ and
\begin{equation}
\rm  h(k_{\bot g}) = \sqrt{2e} \frac{k_{\bot g}}{M_1} e^{-k^2_{\bot g} / M^2_1 }.
\end{equation}
Therefore the $\rm k_{\bot g}$ dependent part of the Sivers function can expressed as follows
\begin{equation}
\rm h(k_{\bot g}) \frac{ e^{-k^2_{\bot g} / \langle k^2_{\bot g} \rangle} }{\pi \langle k^2_{\bot g} \rangle}
=\frac{\sqrt{2e}}{\pi} \sqrt{\frac{1-\rho}{\rho}} k_{\bot g} \frac{ e^{-k^2_{\bot g} /\rho \langle k^2_{\bot g} \rangle} }{ \langle k^2_{\bot g} \rangle ^{3/2} },
\end{equation}
where
\begin{equation}
\rm \rho=\frac{M^2_1}{ \langle k^2_{\bot g} \rangle + M^2_1 } .
\end{equation}
Here $\rm N_g, \alpha, \beta, M_1$ are all parameters determined by fits to data and e is Euler's number. The two extractions of the GSF, namely SIDIS1 and SIDIS2 were obtained by fitting to data.
The numerical values of free parameters $\rm a_f$, $\rm b_f$ and $\rm N_f$ have been estimated by global fit of single spin asymmetry in semi-inclusive deep-inelastic scattering (SIDIS) process \cite{Anselmino:2016uie,DAlesio:2015fwo}. However, only the u and d quark's free parameters are extracted \cite{Anselmino:2005ea} and gluon parameters $\rm a_g,b_g$ and $\rm N_g$ are not known yet.
To estimate SSA we use two parameterizations to attain the best fit parameters of gluon Sivers function \cite{Boer:2003tx}
\begin{eqnarray}\label{Ngformula}
\begin{aligned}
\rm (a) \ \mathcal{N}_g (x) &\rm=  \frac{\mathcal{N}_u(x)+\mathcal{N}_d(x)}{2} \\
\rm (b) \ \mathcal{N}_g (x) &\rm=  \mathcal{N}_d (x)
\end{aligned}
\end{eqnarray}
The best fit parameters are tabularized in the following section.

The above simplified expression that we adopt, both for the unpolarized distribution and the Sivers function, is known as the Gaussian factorization ansatz. It has been favorably checked against the data in the Drell-Yan \cite{DAlesio:2004eso} and SIDIS \cite{Schweitzer:2010tt}. However, it is still far less clear whether it is a suitable method to study DGLAP evolution of TMDs. The factorization ansatz which is assumed to hold at an initial condition scale, is broken at higher scales and the breaking increases with the evolution range and with decreasing x \cite{Broniowski:2017gfp}. In particular it is completely broken in the very low x limit \cite{Yao:2018vcg}. As it was found that the DGLAP evolution approach could also not be able to describe the Z-boson high transverse momentum distribution in Dree-Yan process at CDF \cite{Melis:2014pna}. Nevertheless, to also explain high $\rm P_T$ data, one has to consider TMD evolution approach which we will study in the following subsection.

The final expressions of the asymmetry can be written in the DGLAP evolution formalism. In considering the $\rm \sin(\phi_3-\phi_s)$ weighted factor, the numerator and the denominator terms of Eq.(\ref{AN}) are given by
\begin{eqnarray} \nonumber
\rm
\frac{d\sigma^\uparrow}{d^2 \bold{p}_{3}} -  \frac{d\sigma^\downarrow}{d^2 \bold{p}_{3}}  &=&\rm
\int dx_\gamma d^2{\bold{k}}_{\bot\gamma}
f_{\gamma/h}(x_\gamma, \bold{k}_{\bot\gamma})  f_{g/p}(x_g,\mu)  \\\nonumber
&&\rm  \times2 N_g(x)
\frac{\sqrt{2e}}{\pi} \sqrt{\frac{1-\rho}{\rho}} k_{\bot g} \frac{ e^{-k^2_{\bot g} /\rho \langle k^2_{\bot g} \rangle} }{ \langle k^2_{\bot g} \rangle ^{3/2} }
\frac{\pi}{s^2 x_g x^2_\gamma} |\overline{\mathcal{M}}|^2 \sin(\phi_{\bot g}-\phi_s)\sin(\phi_3-\phi_s)  \\
\rm
\frac{d\sigma^\uparrow}{d^2 \bold{p}_{3}} +  \frac{d\sigma^\downarrow}{d^2 \bold{p}_{3}}  &=& \rm
2 \int dx_\gamma d^2{\bold{k}}_{\bot\gamma} f_{\gamma/h}(x_\gamma, \bold{k}_{\bot\gamma})
f_{g/p}(x_g,\mu)\frac{1}{\pi \langle k^2_{\bot g} \rangle } e^{-k^2_{\bot g} / \langle k^2_{\bot g} \rangle }
\frac{\pi}{s^2 x_g x^2_\gamma} |\overline{\mathcal{M}}|^2 \ .
\end{eqnarray}

\subsection{Sivers asymmetry and parameterization in TMD evolution}

Here, we study the TMD evolution approach. Since TMDs depend on varoius energy scales, the TMD pdf $\rm f(x, k_\bot, Q)$ is best described through its Fourier transform into coordinate space (an impact parameter $\rm b_\bot$-space) which is given by
\begin{eqnarray}
\rm f(x,b_\bot,Q) = \int d^2 \bold{k}_\bot e^{-i \bold{k}_\bot\cdot\bold{b}_\bot} f(x,k_\bot,Q)
\end{eqnarray}
with the inverse Fourier transformation
\begin{eqnarray}
\rm f(x,k_\bot,Q) = \frac{1}{(2\pi)^2} \int d^2 \bold{b}_\bot e^{i \bold{k}_\bot\cdot\bold{b}_\bot } f(x,b_\bot,Q) .
\end{eqnarray}
The evolution of $\rm b_\bot$-space TMD pdfs can then be written as
\begin{eqnarray}
\rm f(x,b_\bot,Q_f) = f(x,b_\bot,Q_i) \times R_{P}(Q_f,Q_i,b_*) \times R_{NP}(Q_f,Q_i,b_\bot)
\end{eqnarray}
where $\rm R_{P}$ is the perturbatively calculable part of the evolution kernel in small $\rm b_{\bot}$ region, $\rm R_{NP}$ is a nonperturbative Sudakov factor in the large $\rm b_{\bot}$ region probably obtained from the experimental data  \cite{Landry:2002ix,Konychev:2005iy}. To combine these regions, a matching procedure is introduced with a parameter $\rm b_{\bot max}$ serving as the boundary between the two regions. There were several different prescriptions \cite{Collins:2016hqq,Bacchetta:2017gcc} in literature. Here we adopt the original Collins-Soper-Sterman (CSS) prescription \cite{Collins:1984kg,Qiu:2000ga,Qiu:2000hf}
\begin{eqnarray}
 \rm b_*=b_\bot/\sqrt{1+(b_\bot/b_{\bot max})^2}, \ \ b_{\bot max}< 1/\Lambda_{QCD}
\end{eqnarray}
which allows a smooth transition from perturbative to nonperturbative regions and avoids the Landau pole singularity in $\rm \alpha_s(\mu_{b_{\bot}})$. The typical value of $\rm b_{\bot max}$ is chosen around 1 $\rm GeV^{-1}$ to guarantee that $\rm b_{*}$ is always in the perturbative region.

In the small $\rm b_{\bot}$ region, the TMD distributions at fixed energy can be expressed as the convolution of the perturbatively calculable coefficients and the corresponding collinear PDFs or the multiparton correlation functions. Following refs \cite{Echevarria:2014xaa,Bacchetta:2013pqa},
we choose an initial scale $\rm Q_i=c/b_*$ to start the TMD evolution, where $\rm c=2 e^{-\gamma_{E}}$ and $\gamma_E\approx 0.577$ is the Euler-Mascheroni constant. Setting $\rm Q_i=c/b_*$ and $\rm Q_f=Q$, the perturbative evolution kernel is given by \cite{Kang:2011mr,Echevarria:2012pw,Echevarria:2014xaa,Aybat:2011ge,Boer:2014tka,Catani:1988vd,Kauffman:1991cx}
\begin{eqnarray}
\rm R_{P}(Q_f,Q_i,b_*)= exp \{ -\int^{Q_f}_{c/b_*} \frac{d\mu^\prime}{\mu^\prime}
\left( A(\alpha_s(\mu^\prime)) \ln \left(\frac{Q^2_f}{\mu^{\prime2}} \right) +B(\alpha_s(\mu^\prime)) \right) \} \times \left(\frac{Q^2_f}{Q^2_i}\right)^{-D(b;Q_i)}
\end{eqnarray}
where $\rm A=\Gamma_{cusp}$ and $\rm B=\gamma^{V}$ with $\rm \frac{dD}{d\log\mu}=\Gamma_{cusp}$. $\rm \Gamma_{cusp}$ and $\rm \gamma^{V}$ are anomalous dimensions and can be expanded perturbative as the series of $\rm \alpha_s/\pi$
\begin{eqnarray} \nonumber
\rm A&=&\rm \sum^{\infty}_{n=1} \left( \frac{\alpha_s}{\pi} \right)^n A_n ,  \\\nonumber
\rm B&=&\rm \sum^{\infty}_{n=1} \left( \frac{\alpha_s}{\pi} \right)^n B_n \ ,\\
\rm D&=&\rm \sum^{\infty}_{n=1} \left( \frac{\alpha_s}{\pi} \right)^n D_n \ .
\end{eqnarray}
The expansion coefficients with the appropriate gluon anomalous dimensions up to the accuracy of next-to-leading-logarithmic (NLL) order are \cite{Idilbi:2006dg,Echevarria:2012pw,deFlorian:2000pr,deFlorian:2001zd,Echevarria:2014xaa}
\begin{eqnarray} \nonumber
\rm A_1 &=&\rm C_A, \\\nonumber
\rm A_2 &=&\rm  \frac{1}{2} C_A \left( C_A (\frac{67}{18} -\frac{\pi^2}{6}) -\frac{10}{9} T_R N_f \right),  \\ \nonumber
\rm B_1 &=&\rm  -\frac{1}{2} \left( \frac{11}{3} C_A -\frac{4}{3} T_R N_f + C_A\delta_{c,8} \right),\ \\
\rm D_1 &=&\rm \frac{C_A}{2} \log\frac{Q^2_i b^2_*}{c^2}
\end{eqnarray}
The Kronecker delta $\rm \delta_{c,8}$ derives from the interference of the initial and final state soft gluon radiation in the color-octet channel ($\rm c=8$) and is absent in the color-singlet channel ($\rm c=1$) \cite{Sun:2012vc}.
The D term vanishes at NLL by choosing the initial scale $\rm Q_i = c/b_*$.

The CSS resummation formalism suggests that the nonperturbative functional is universal. Its role is similar to that of the parton distribution function in any fixed order perturbative calculation and its origin is due to the long distance effects that are incalculable at the present, and its value must be determined from data. The general formula of nonperturbative function is given by
\begin{equation}
  \rm R_{ij}^{NP}(b_\bot,Q,x_{A},x_{B})=\exp [-\ln(Q^{2}/Q_{0}^{2})g_{1}(b_\bot)-g_{i/A}(x_{A},b_\bot)-g_{j/B}(x_{B},b_\bot)]
\end{equation}
where the functions $\rm g_{1}(b_\bot)$, $\rm g_{i/A}(x_{A},b_\bot)$ and $\rm g_{j/B}(x_{B},b_\bot)$ must be extracted from data with the constraint that $\rm R_{ij}^{NP}(0,Q,x_{A},x_{B})=1$. They should go to zero as $\rm b_\bot\rightarrow 0$. $\rm x_{A}$ and $\rm x_{B}$ represent the longitudinal momentum fractions of the incoming hadrons carried by the initial state partons (photon and gluon).
$\rm \ln(Q^{2}/Q_{0}^{2})g_{1}(b_\bot)$ dependence is proposed by the infrared renormalon contributions which is a certain pattern of perturbative expansions related to the small and large momentum behavior \cite{Beneke:1998ui}. Moreover, $\rm g_{1}(b_\bot)$ only depends on $\rm Q$, whereas $\rm g_{i/A}(x_{A},b_\bot)$ and $\rm g_{j/B}(x_{B},b_\bot)$ in general rely on $\rm x_{A}$ or $\rm x_{B}$, and their values can depend on the flavor of the initial state partons. The nonperturbative element of the evolution kernel cannot be evaluated and a parametrized form has to be selected. There are many extractions for the nonperturbative part mentioned in literature inspired by refs \cite{Landry:2002ix,Qiu:2000ga} and widely used to parameterize $\rm R_{ij}^{NP}(b_\bot,Q,x_{A},x_{B})$ for TMD distributions. Some often-used functional forms are defined in four types as follows. 
\begin{itemize}
\item The nonperturbative distribution introduced by Davis, Webber and Stirling (DWS) \cite{Davies:1984sp} is given by
\begin{equation}
  \rm R_{NP}^{DWS}(b_\bot,Q,x_{A},x_{B})= \exp [-b^2_\bot(g_{1}+g_{2}\ln(Q^{2}/2Q_{0}^{2}))]
  \end{equation}
where $\rm g_{1}$ and $\rm g_{2}$ are flavor independent fitting parameters. DWS is a pure Gaussian form. The CSS $\rm b$-space resummation formalism with DWS distribution offers a reasonable description of the Drell-Yan data from Fermilab experiment E288 at $\rm \sqrt{s}=27.4$ GeV \citep{Ito:1980ev}
and CERN ISR experiment R209 at $\rm \sqrt{s}=67$ GeV \citep{Antreasyan:1981uv,Antreasyan:1981eg}.
\item So as to incorporate possible $\rm \ln(x_{A}x_{B})$ dependence which is linear in
$\rm b_\bot$, Landinsky and Lyan (LY) \cite{,Ladinsky:1993zn,Landry:1999an}  suggested a revised functional form for the $ \rm R_{ij}^{NP}$ with extra parameter $\rm g_{3}$. LY is able to fit the R209 Drell-Yan data and CDF data on $\rm W$ and $\rm Z$ production from  Fermilab and is formulated by   \begin{equation}
  \rm  R_{NP}^{LY}(b_\bot,Q,x_{A},x_{B})= \exp [-b^2_\bot(g_{1}+g_{2}\ln(Q^{2}/2Q_{0}^{2}))+b_\bot g_{1}g_{3}\ln(100x_{A}x_{B})].
  \end{equation}
LY is not a pure Gaussian form.
\item Landry, Brock, Landinsky and Yuan (LBLY) \cite{Landry:1999an,Melis:2014pna} perfomed a much more extensive global fit to the low energy Drell-Yan data along with high energy $\rm W$ and $\rm Z$ data by using both DWS and LY parametrizations. Its expression is stated as 
  \begin{equation}
  \rm  R_{NP}^{LBLY}(b_\bot,Q,x_{A},x_{B})= \exp [-b^2_\bot(g_{1}+g_{2}\ln(Q^{2}/2Q_{0}^{2})+g_{1}g_{3}\ln(100x_{A}x_{B}))].
  \end{equation}
  LBLY is also a pure Gaussian form.
\item Recently, the nonperturbative form factor $\rm R_{ij}^{NP}$ of BLNY associated with the unpolarized TMD PDF of the proton has been simplified. The updated BLNYs (UBLNY) \cite{Aybat:2011zv,Sun:2012vc,Collins:1985xx,Su:2014wpa,Melis:2014pna} are constructed and fitted such as to describe the low energy SIDIS as well as high energy Drell-Yan and $\rm Z$ production data. They can establish the universality property of the TMD distributions between DIS and Drell-Yan process \cite{Su:2014wpa}. The UBLNY in ref.\cite{Melis:2014pna} has been chosen and used in \cite{Su:2014wpa} to study the unpolarized pp Drell-Yan process,
\begin{eqnarray}\label{NNupdate}
\rm R_{NP}^{UBLNY}(b_\bot,Q,x_{A},x_{B})=\exp \{ -[ g_{1}b^2_\bot + g_{2}\ln\frac{b_\bot}{b_*}\ln\frac{Q}{Q_{0}}
+ g_{3} b^2_\bot ( (\frac{x_0}{x_{A}})^\lambda + (\frac{x_0}{x_{B}})^\lambda ) ]  \}.
\end{eqnarray}
With the parameterization in Table \ref{table1}, the Eq.(\ref{NNupdate}) has been reduced to  
  \begin{eqnarray}
  \rm R_{NP}^{UBLNY}(b_\bot,Q,x_{A},x_{B}) = exp \{ -[ \frac{g_1}{2} b^2_\bot + \frac{g_2}{2} \ln\frac{b_\bot}{b_*} \ln\frac{Q}{Q_0} ] \}
\end{eqnarray}
which has been used for all quark TMDPDFs. In case of gluon TMDPDFs, $\rm g_2$ is to be multiplied by a factor of $\rm C_A/C_F$. In comparison to the quark parametrization, the coefficient of the term proportional to $\rm \ln(Q)$ is enhanced by a color factor while the the intrinsic part is kept unchanged \cite{Kang:2017glf}. 
\end{itemize}
\begin{table}[htpb]
	\setlength{\tabcolsep}{5mm}
	\begin{center}
		\begin{tabular}{|c|c|c|c|c|c|c|c|}
			\hline
			  $\rm R_{NP}$& $\rm g_{1}/GeV^{2}$ & $\rm g_{2}/GeV^{2}$ & $\rm g_{3}/GeV^{2}$ & $\rm Q_{0}/GeV $ & $\rm b_{max}/GeV^{-1}$ & $\rm x_{0}$& $\rm \lambda$ \\
			\hline
			DWS & 0.15 & 0.4 &   & 2&0.5 & &    \\
			\hline
			LY & 0.11 & 0.58 & -1.5 & 1.6  & 0.5& &  \\
			\hline
		BLNPY & 0.21 & 0.68 & -0.12 &1.6  & 0.5 & &   \\
			\hline
		UBLNPY & 0.212 & 0.84 & 0.0 &1.5  & 1.5& 0.01 & 0.2  \\
			\hline
		\end{tabular}
	\end{center}
\vspace{-0.5cm}
	\caption{\label{table1}Best fit parameters of nonperturbative Sudakov factor, $\rm R_{NP}$.}
\end{table}

Combining the former discussions and following Ref.\cite{Echevarria:2014xaa}, one can expand the TMD $\rm f(x, b_{\bot}\ Q)$ at the initial scale in terms of its corresponding collinear function and keep only the leading order term, which is just the collinear PDF. The TMD evolution equation of the unpolarized gluon TMD-PDF in terms of collinear PDF in $\rm b_{\bot}-$space is finally given by
\begin{eqnarray}
\rm f_{g/p}(x_g,b_\bot,Q) = f_{g/p} (x_g,c/b_*) \times exp \{ -\int^{Q}_{c/b_*} \frac{d\mu'}{\mu'} \left( A ln\left(\frac{Q^2}{\mu^{'2}} \right) +B \right) \}
exp \{ -[ \frac{g_1}{2} b^2_\bot + \frac{g_2}{2} \ln\frac{b_\bot}{b_*} \ln\frac{Q}{Q_0} ] \}.
\end{eqnarray}

For the gluon sivers function, its azimuth-dependent part (in b-space) in the so-called Trento convention \cite{Aybat:2011ge} is
\begin{eqnarray}
\rm f^{\bot g(\alpha)}_{1T} (x_g,b_\bot,Q) = \frac{1}{m_p} \int d^2 \bold{k}_\bot e^{-i \bold{k}_\bot \cdot \bold{b}} k^\alpha_{\bot} f^{\bot g}_{1T}(x_g,k^2_\bot,Q).
\end{eqnarray}
Expanding this in $\rm b_\bot$ and keeping the leading term we get
\begin{eqnarray} \label{F1Tg}
\rm f^{\bot g(\alpha)}_{1T} (x_g,b_\bot,Q) \simeq -\frac{ib_\bot^\alpha}{2 m_p} \int d^2 \bold{k}_\bot |k_\bot|^2 f^{\bot g}_{1T}(x,k^2_\bot,Q) = \frac{ib_\bot^\alpha}{2} T_{g,F}(x_g,x_g,Q).
\end{eqnarray}
Here $\rm T_{g,F}(x_g,x_g,Q)$ \cite{Kang:2011mr,Kang:2011hk,Kouvaris:2006zy} is the twist-3 Qiu-Sterman quark-gluon correlation function, treated at the leading order as a sivers function. It is the first $\rm k_T$ moment term of the sivers function and plays a significant role in the theoretical description of transverse SSA  in the framework of collinear factorization. Qiu-Sterman functions can also  determine the large transverse momentum tail of gluon Sivers function. Considering Eq.(\ref{F1Tg}) and the derivative of the Sivers function in b-space, we thus get
\begin{eqnarray}
\rm f^{'\bot g}_{1T} (x_g,b_\bot,Q) = \frac{\partial f^{\bot g}_{1T} (x_g,b_\bot) }{\partial b_\bot} = -i \frac{m_p b_\bot}{b_\bot^\alpha} f^{\bot g(\alpha)}_{1T} (x_g,b_\bot,Q)
\simeq \frac{m_p b_\bot}{2} T_{g,F} (x_g,x_g,Q),
\end{eqnarray}
which satisfies the same evolution equation for the perturbative part as the unpolarized TMD PDF. For the nonperturbative part, we follow ref.\cite{Echevarria:2014xaa}, where the authors proposed a Sudakov form factor in the evolution formalism, which can lead to a good description of the transverse momentum distribution for different processes such as SIDIS, DY dilepton and W/Z boson production in pp collisions. The nonperturbative Sudakov form factor $\rm S_{NP}$ for the Sivers function has the form
\begin{eqnarray}
\rm R_{NP} = exp \{ -b^2_\bot ( g^{Sivers}_1 + \frac{g_2}{2}\ln\frac{Q}{Q_0} ) \}
\end{eqnarray}
where the parameter $\rm g^{Sivers}_1$ relevant to the averaged intrinsic transverse momenta squared $\rm g^{Sivers}_1=\langle k^2_{\bot s} \rangle_{Q_0}/4 =0.071\ GeV^2$, $\rm g_2$ is universal for all different types of TMDs, spin-independent \cite{Echevarria:2014xaa} and equal to $\rm \frac{1}{2} g_2 = 0.08\ GeV^2$, and here $\rm Q_0=\sqrt{2.4}\ GeV$ and $\rm b_{max}=1.5\ GeV^{-1}$. So that, in the case of the Sivers function, the evolution of its derivative can be written in the form of
\begin{eqnarray}
\rm
f'^{\bot g}_{1T}(x_g,b_\bot,Q_f) = f'^{\bot g}_{1T}(x_g,b_\bot,Q_i)
exp \{ -\int^{Q_f}_{Q_i} \frac{d\mu'}{\mu'} \left( A ln\left(\frac{Q^2_f}{\mu^{'2}} \right) + B \right) \}
exp \{ -b^2_\bot ( g^{Sivers}_1 + \frac{g_2}{2}  \ln\frac{Q_f}{Q_0} ) \} .
\end{eqnarray}
Setting the initial scale equal $\rm Q_i=c/b_*$ and $\rm Q_f=Q$, we finally have
\begin{eqnarray}
\rm
f'^{\bot g}_{1T}(x_g,b_\bot,Q) = \frac{m_p b_{\bot}}{2} T_{g,F}(x_g,x_g,c/b_*)
exp \{ -\int^{Q}_{c/b_*} \frac{d\mu'}{\mu'} \left( A ln\left(\frac{Q^2}{\mu^{'2}} \right) + B \right) \}
exp \{ -b^2_\bot ( g^{Sivers}_1 + \frac{g_2}{2} \ln\frac{Q}{Q_0} ) \} .
\end{eqnarray}
Here the Qiu-Sterman function $\rm T_{g,F}(x_g,x_g,Q)$ can be parameterized proportionally to the collinear PDF as
\begin{eqnarray}
 \rm T_{g,F}(x_g,x_g,Q) = N_g(x_g) f_{g/p}(x_g,Q)
\end{eqnarray}
with $\rm N_g(x_g)$ defined in Eq.(\ref{Ng}). 

Therefore, the expressions for the TMDs in $\rm k_{\bot}$-space can be obtained by Fourier transforming the $\rm b_{\bot}$-space expressions
\begin{eqnarray} \nonumber
\rm f_{g/p}(x_g,k_{\bot g},Q) &=&\rm \frac{1}{2\pi} \int^{\infty}_0 db_{\bot} b_{\bot} J_0(k_{\bot g}b_{\bot}) f_{g/p}(x_g,b_{\bot},Q) \\
\rm f^{\bot g}_{1T}(x_g,k_{\bot g},Q) &=&\rm \frac{-1}{2\pi k_{\bot g}} \int^{\infty}_0 db_{\bot} b_{\bot} J_1(k_{\bot g} b_{\bot}) f^{'\bot g}_{1T}(x_g,b_{\bot},Q)
\end{eqnarray}
where $\rm J_{0/1}$ are the zeroth/1st order Bessel functions of the first kind. Using the above expressions, the asymmetry including the weighted factors $\rm \sin(\phi_3-\phi_s)$ can be written in the TMD evolution framework as follows
\begin{eqnarray}\nonumber
\rm
\frac{d\sigma^\uparrow}{d^2 \bold{p}_{3}} - \frac{d\sigma^\downarrow}{d^2 \bold{p}_{3}}  &=&\rm
\int dx_\gamma d^2{\bold{k}}_{\bot\gamma} f_{\gamma/h}(x_\gamma, \bold{k}_{\bot\gamma})   \\\nonumber
&&\rm  \frac{-1}{2\pi k_{\bot g}} \int^{\infty}_0 db_\bot b_\bot J_1(k_{\bot g} b_\bot) f^{'\bot g}_{1T}(x,b_\bot, \mu)
\frac{-2k_{\bot g}}{m_P}
\frac{\pi}{s^2 x_g x^2_\gamma} |\overline{\mathcal{M}}|^2 \sin(\phi_{\bot g}-\phi_s) \sin(\phi_3-\phi_s)  \\
\rm
\frac{d\sigma^\uparrow}{d^2 \bold{p}_{3}} + \frac{d\sigma^\downarrow}{d^2 \bold{p}_{3}}  &=&\rm
2 \int dx_\gamma d^2{\bold{k}}_{\bot\gamma} f_{\gamma/h}(x_\gamma, \bold{k}_{\bot\gamma})
\frac{1}{2\pi}\int^{\infty}_0 db_{\bot}b_{\bot} J_0(k_{\bot g}b_\bot) f_{g/p}(x_g,b_\bot, \mu)
\frac{\pi}{s^2 x_g x^2_\gamma} |\overline{\mathcal{M}}|^2 .
\end{eqnarray}

\section{NUMERICAL RESULTS}
\label{results}

In the following section, we discuss the numerical results of the photoproduction of $\rm J/\psi$ by using some physical parameters such as: $\rm m_{p}=0.94$ GeV as the mass of proton, and $\rm Q_{\max }^{2}$=2 GeV are taken. The mass of the heavy quark is chosen as $\rm m_{c}$=1.548 GeV. The mass of $\rm J/\psi $ is literally put at $\rm M=2m_{c}$. The colliding energies used in this paper are $\rm \sqrt{s}$ = 115 GeV (AFTER@LHC), $\rm \sqrt{s}$ = 200 GeV (RHIC1) and $\rm \sqrt{s}$ = 500 GeV (RHIC2). CTEQ6L1 \cite{Pumplin:2002vw} is used for the PDF which is probed at the factorization scale chosen as $\rm \mu_{f}=M_{T}$, where $\rm m_{T}=\sqrt{\left( p_{T}^{\mathcal{Q}}\right)^{2}+m_{\mathcal{Q}}^{2}}$ is the $\rm m_\mathcal{Q}$ transverse mass. The numerical values of the best fit parameters of nonperturbative Sudakov factor are given in Table \ref{table1}. The numerical values of best fit parameters for DGLAP and TMD evolutions \cite{Echevarria:2014xaa,Mukherjee:2016qxa,DAlesio:2015fwo,Anselmino:2016uie} at $\rm Q_0=\sqrt{2.4}$ GeV are listed in Table \ref{table2}. The numerical evaluation of the Sudakov factor in large impact parameter region at low transverse momentum is handled by the introduction of a nonperturbative function in the CSS resummation formalism. Numerical calculations are carried out by in-house monte carlo generator.
\begin{table}[htpb]
	\setlength{\tabcolsep}{5mm}
	\begin{center}
		\begin{tabular}{|c|c|c|c|c|c|c|c|}
			\hline
			  & $\rm N_a$ & $\alpha$ & $\beta$ & $\rm M_1^2$ GeV$^2$ & $\rho$ & $\rm \langle k_\perp^2 \rangle$ GeV$^2$ & Notation \\
			\hline
			g & 0.65 & 2.8 & 2.8 &  & 0.687 & 0.25 & SIDIS1 \\
			\hline
			g & 0.05 & 0.8 & 1.4 &  & 0.576 & 0.25 & SIDIS2 \\
			\hline
			u & 0.18 & 1.0 & 6.6 & 0.8 &  & 0.57 & BV-a \\
			\hline
			d & -0.52 & 1.9 & 10.0 & 0.8 &  & 0.57 & BV-b \\
			\hline
			u & 0.106 & 1.051 & 4.857 &  &  & 0.38 & TMD-a \\
			\hline
			d & -0.163 & 1.552 & 4.857 &  &  & 0.38 & TMD-b \\
			\hline
		\end{tabular}
	\end{center}
\vspace{-0.5cm}
	\caption{\label{table2}Best fit parameters of Sivers function.
}
\end{table}
From Eq.(\ref{Ngformula}), we symbolize the parametrization (a) and (b) as TMD-a and TMD-b respectively. The choice of the LDMEs for $\rm J/\psi $ is taken from \cite{Chao:2012iv,Mukherjee:2016cjw} as shown in Table \ref{table3}.
\begin{table}[htpb]
\centering
\begin{tabular}{lll}
\hline
$\rm \langle 0|\mathcal{O}_{1,8}^{J/\psi }(^{2S+1}L_{J})|0\rangle$      & Set-I   & Set-II \\
\hline
$\rm \langle 0|\mathcal{O}_{8}^{J/\psi }(^{1}S_{0})|0\rangle$/GeV$^3$       &  $8.9\times 10^{-2}$   & $9.7\times 10^{-2}$  \\
$\rm \langle 0|\mathcal{O}_{8}^{J/\psi }(^{3}P_{0})|0\rangle$/GeV$^5$      & $1.26\times 10^{-2}$    & -$2.14\times 10^{-2}$   \\
\hline
\end{tabular}
\vspace{-0.2cm}
\caption{\label{table3}Numerical values of LDME.}
\end{table}
For $\rm \langle 0|\mathcal{O}_{8}^{J/\psi }(^{3}P_{J})|0\rangle $ with $\rm J$=1,2, and following the heavy-quark spin symmetry, we get the relations:
\begin{equation}
\rm \langle 0|\mathcal{O}_{8}^{J/\psi }(^{3}P_{J})|0\rangle =(2J+1)\langle 0|\mathcal{O}_{8}^{J/\psi}(^{3}P_{0})|0\rangle.
\end{equation}

\begin{figure}[htp]
\centering
   \includegraphics[height=4.6cm,width=4.4cm]{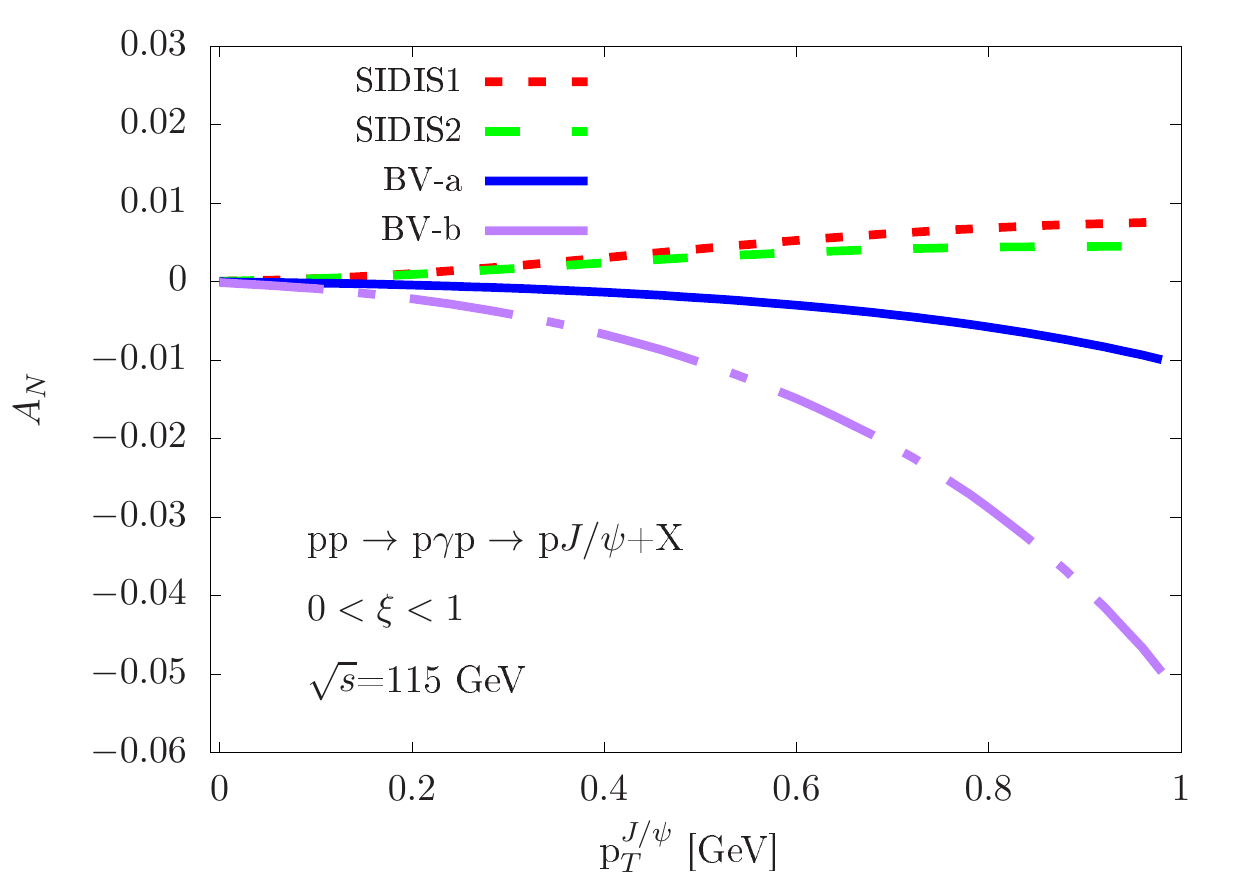}
   \includegraphics[height=4.6cm,width=4.4cm]{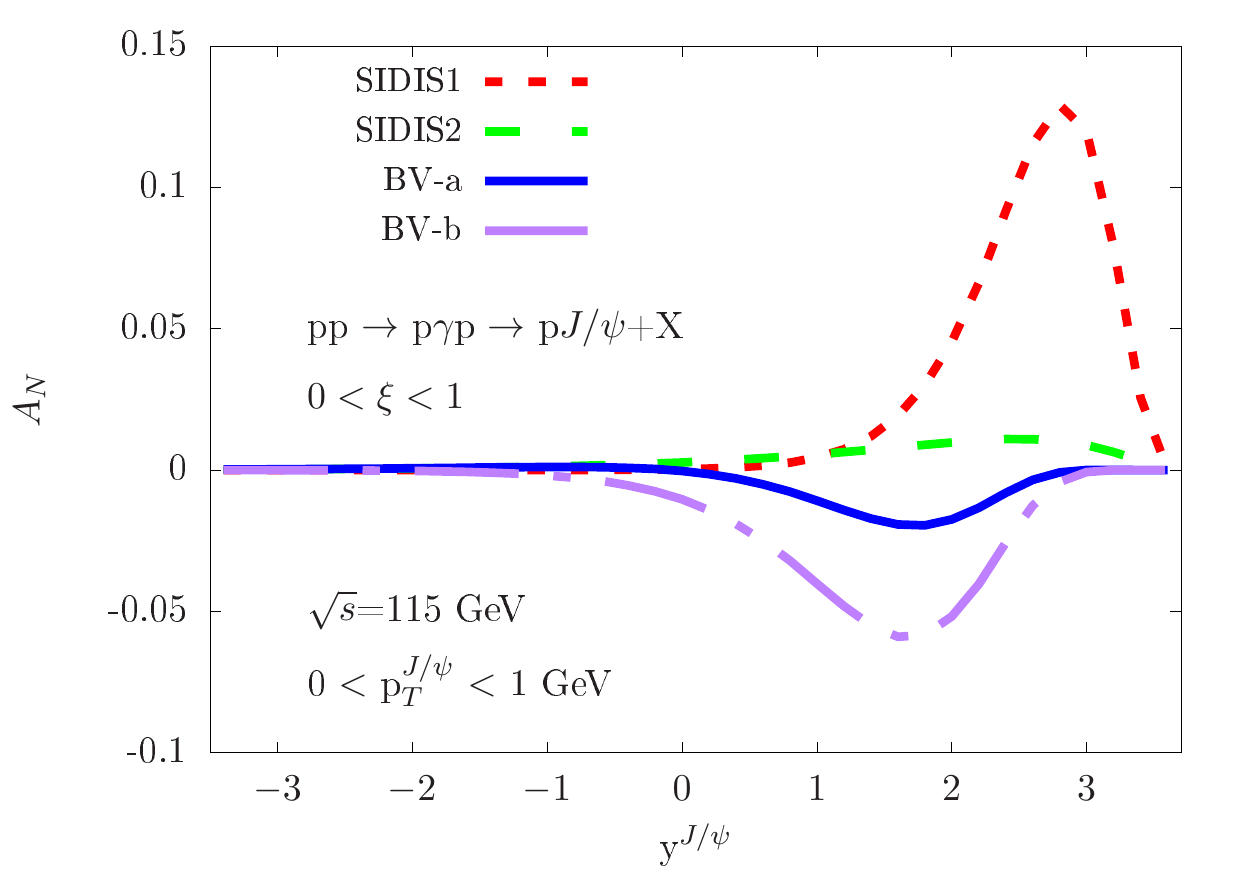}
   \includegraphics[height=4.6cm,width=4.4cm]{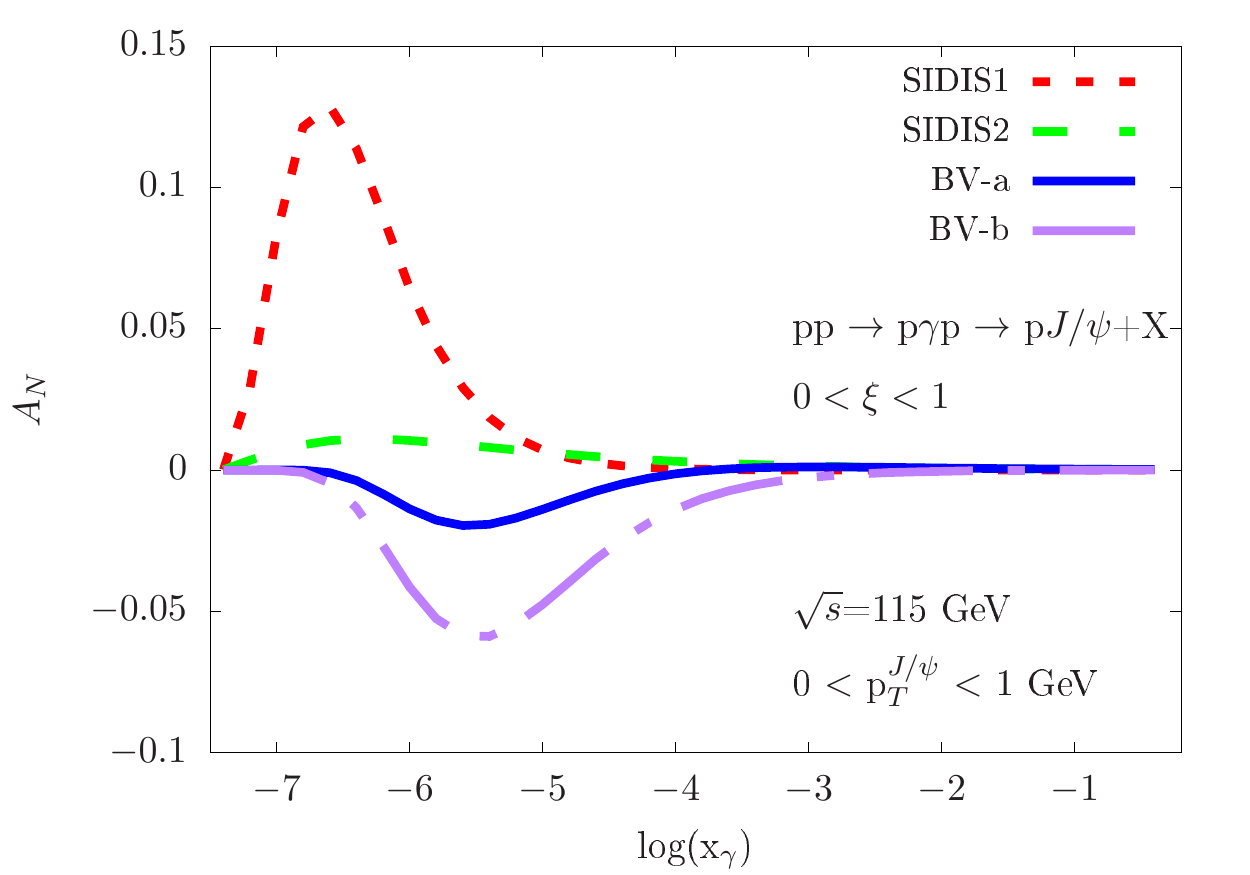}
   \includegraphics[height=4.6cm,width=4.4cm]{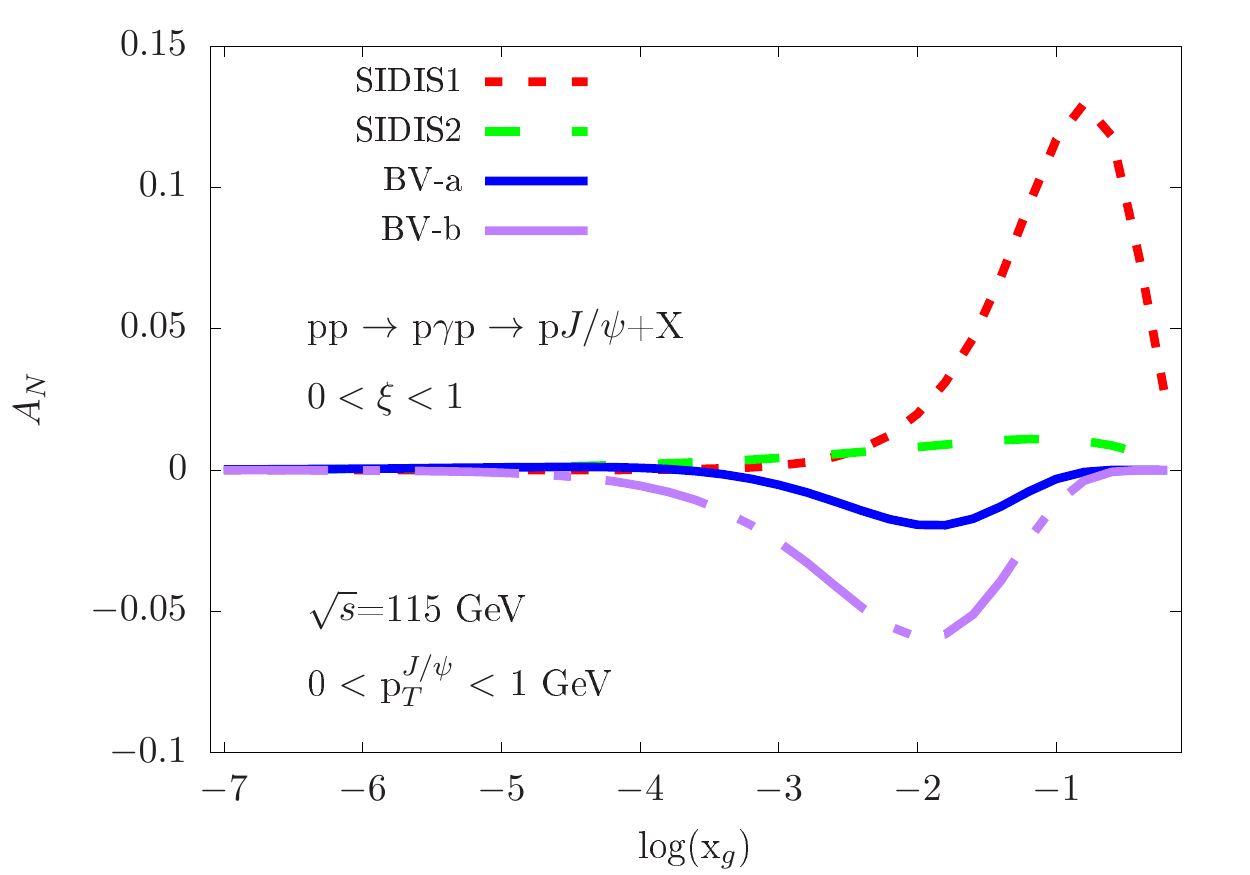}
   \includegraphics[height=4.6cm,width=4.4cm]{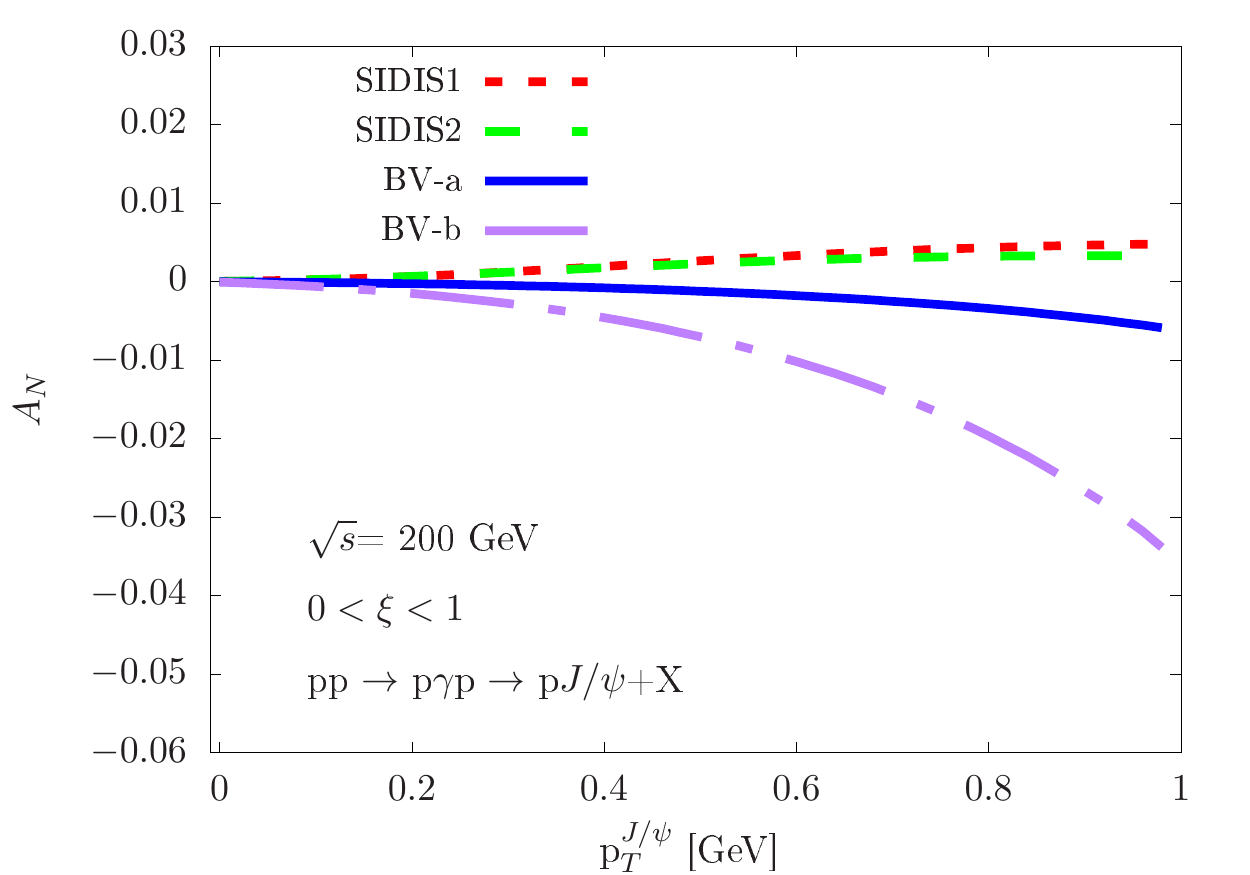}
   \includegraphics[height=4.6cm,width=4.4cm]{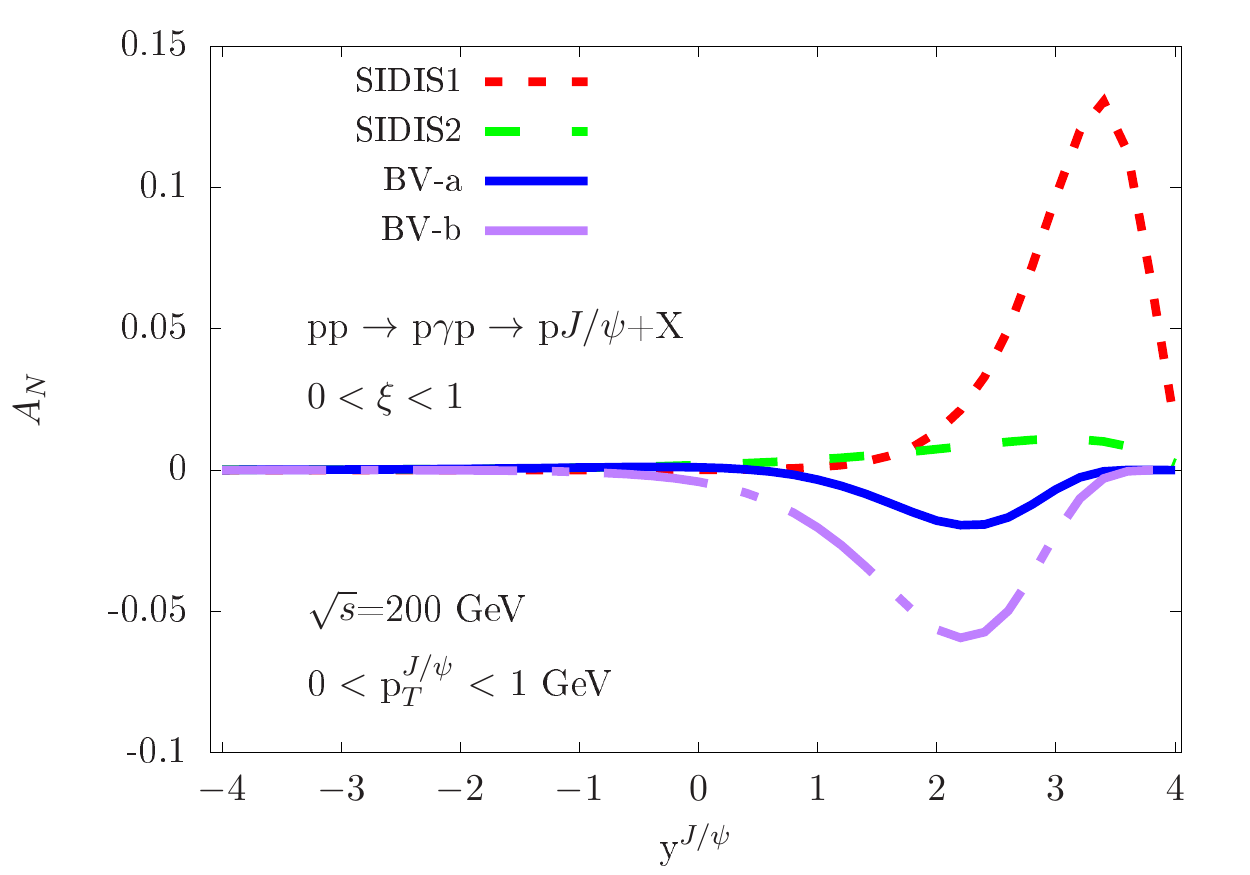}
   \includegraphics[height=4.6cm,width=4.4cm]{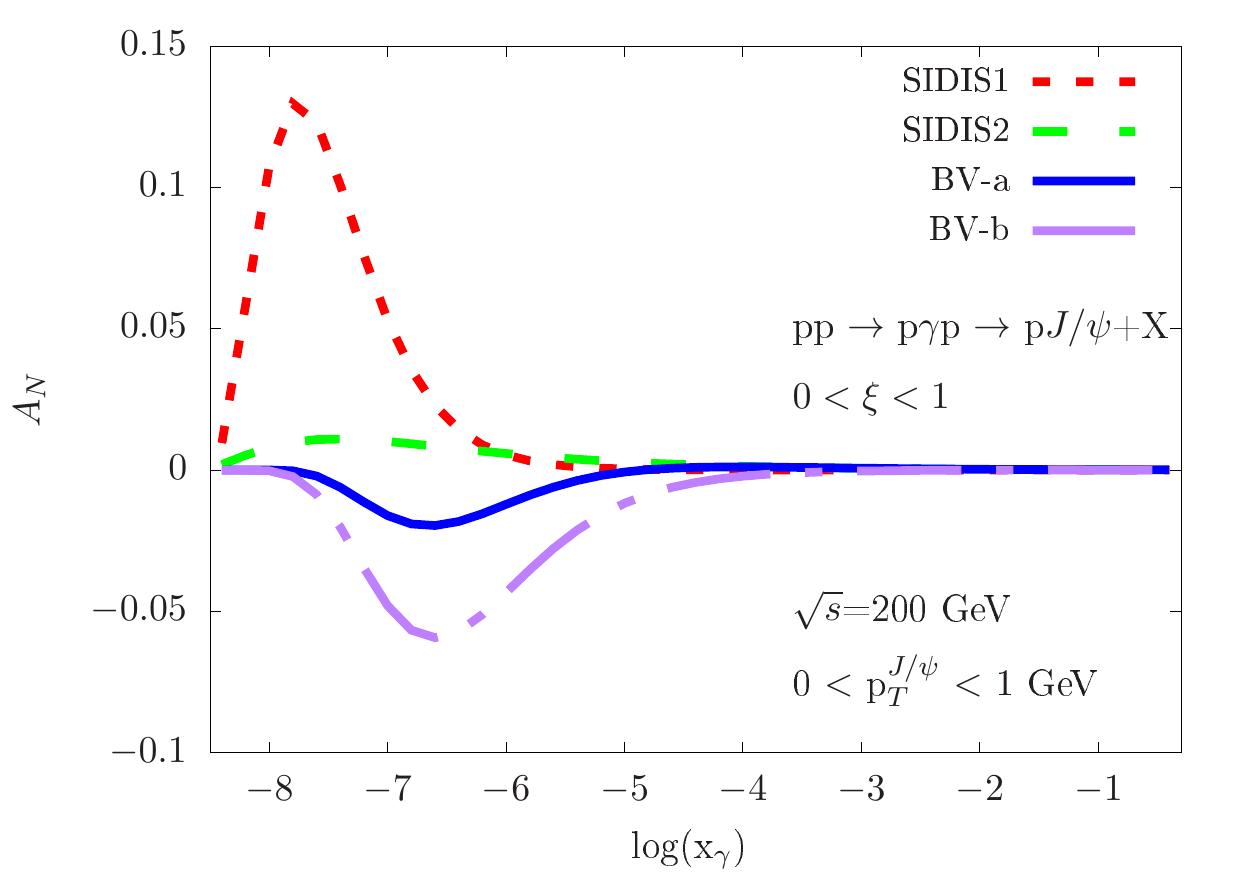}
   \includegraphics[height=4.6cm,width=4.4cm]{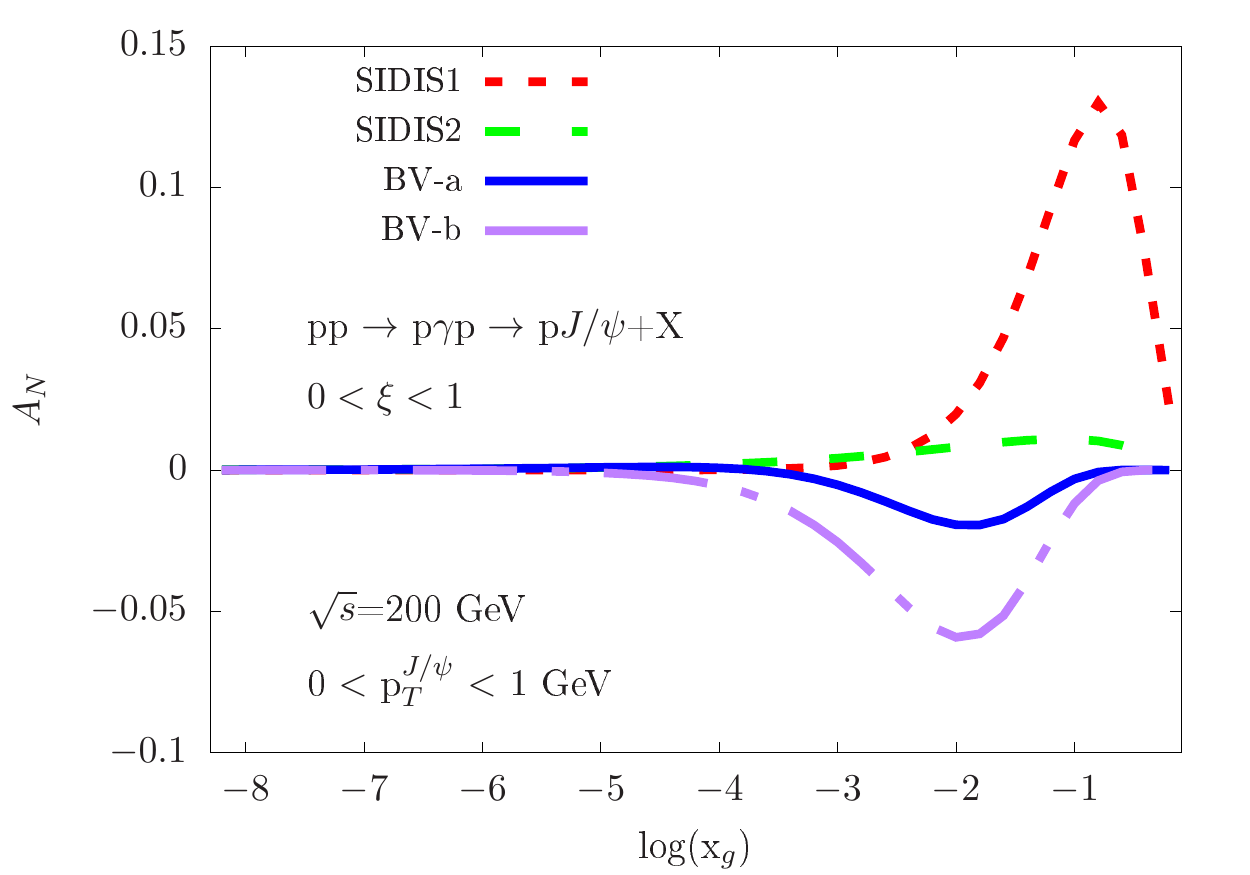}
   \includegraphics[height=4.6cm,width=4.4cm]{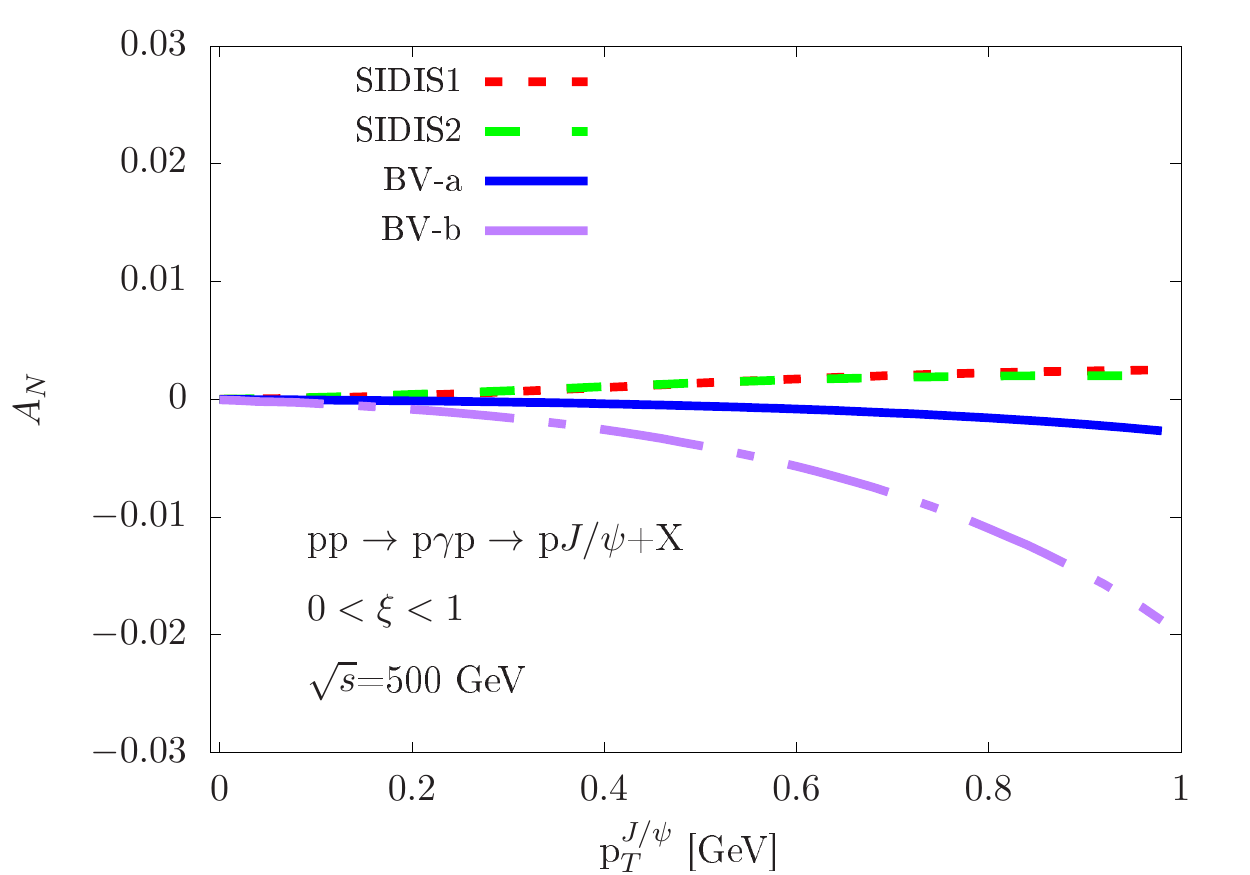}
   \includegraphics[height=4.6cm,width=4.4cm]{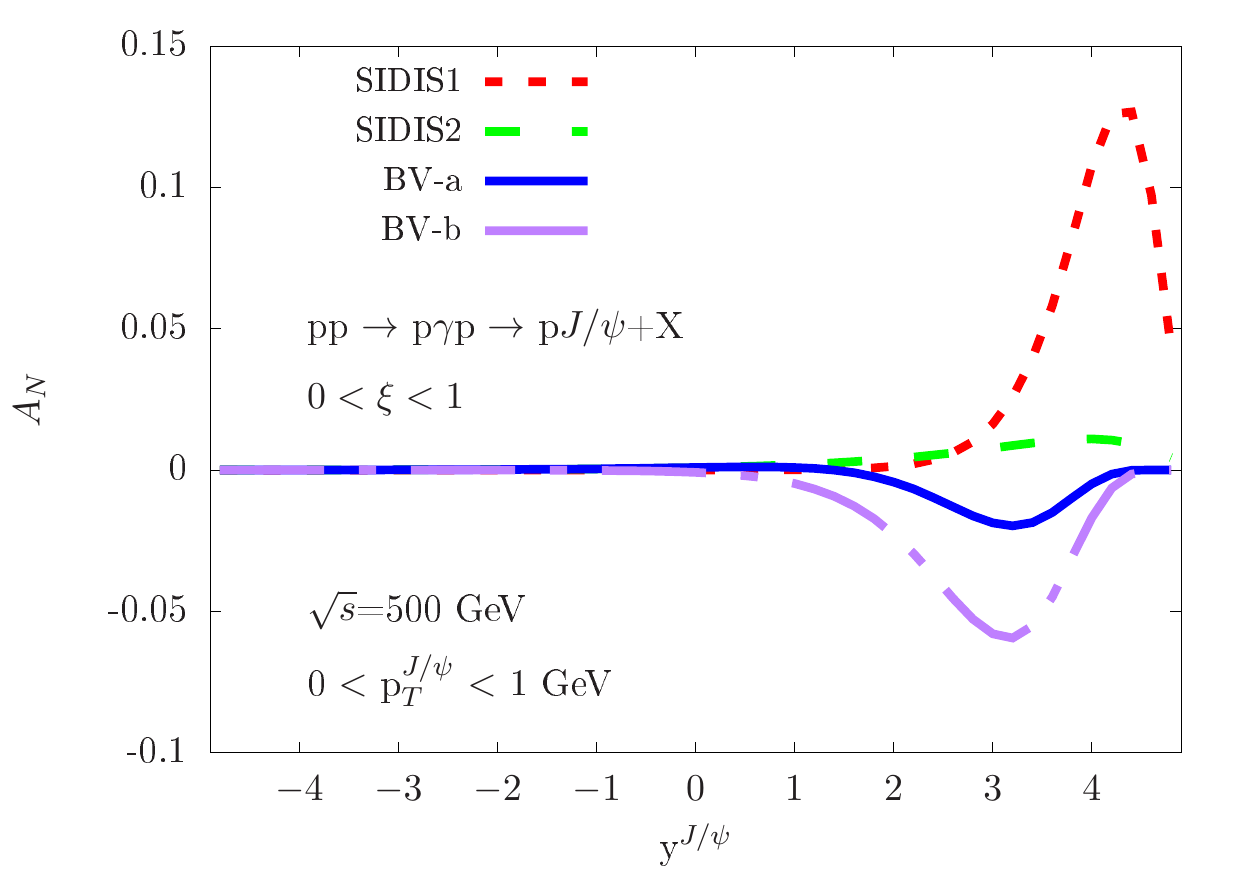}
   \includegraphics[height=4.6cm,width=4.4cm]{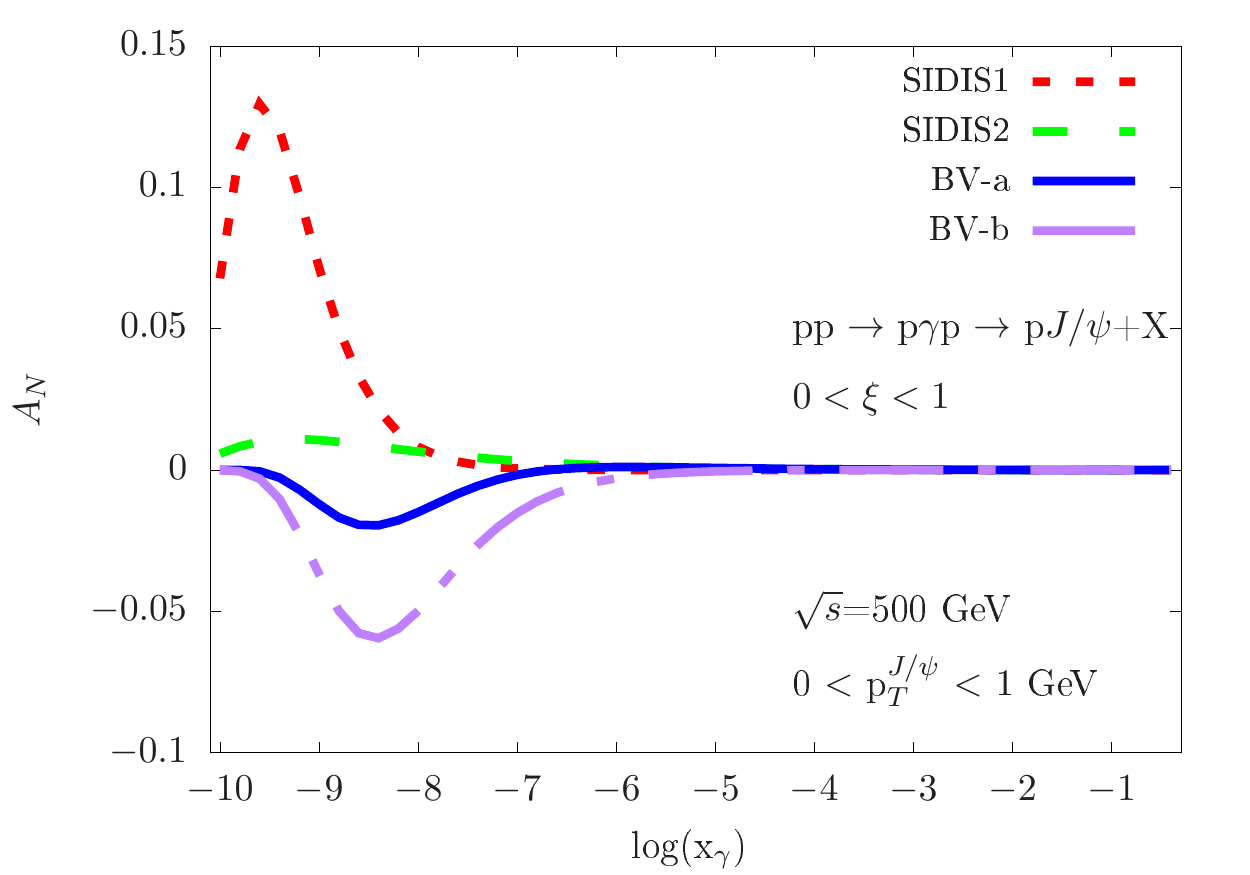}
   \includegraphics[height=4.6cm,width=4.4cm]{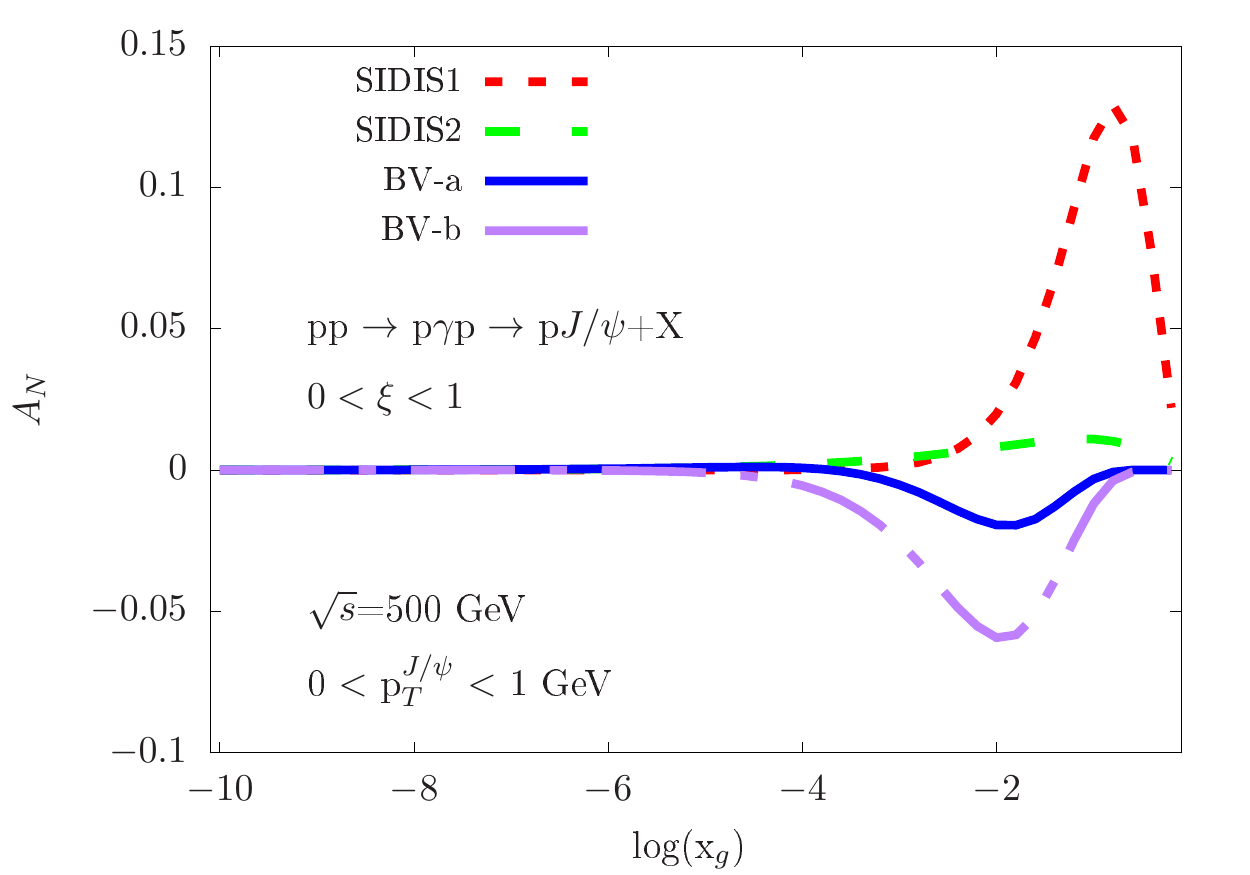}
   \caption{ \normalsize (color online)
Single spin asymmetry in $\rm pp^{\uparrow}\to p\gamma p^{\uparrow}\to p\mathcal{Q}+X$ process
as a function of  $\rm p_{T}^{J/\psi}$ (left column panels), $\rm y^{J/\psi}$ (left middle column  panels), $\rm log{(x_{\gamma})}$ (right middle column panels) and $\rm log{(x_{g})}$ (right column panels) at $\rm \sqrt{s}$ = 115 GeV (AFTER@LHC), $\rm \sqrt{s}$ = 200 GeV (RHIC1) and $\rm \sqrt{s}$ = 500 GeV (RHIC2) using DGLAP ( SIDIS1, SIDIS2, BV-a and BV-b).}
\label{fig4:115GeV}
\end{figure}

\begin{figure}[htp]
\centering
   \includegraphics[height=4.8cm,width=5.0cm]{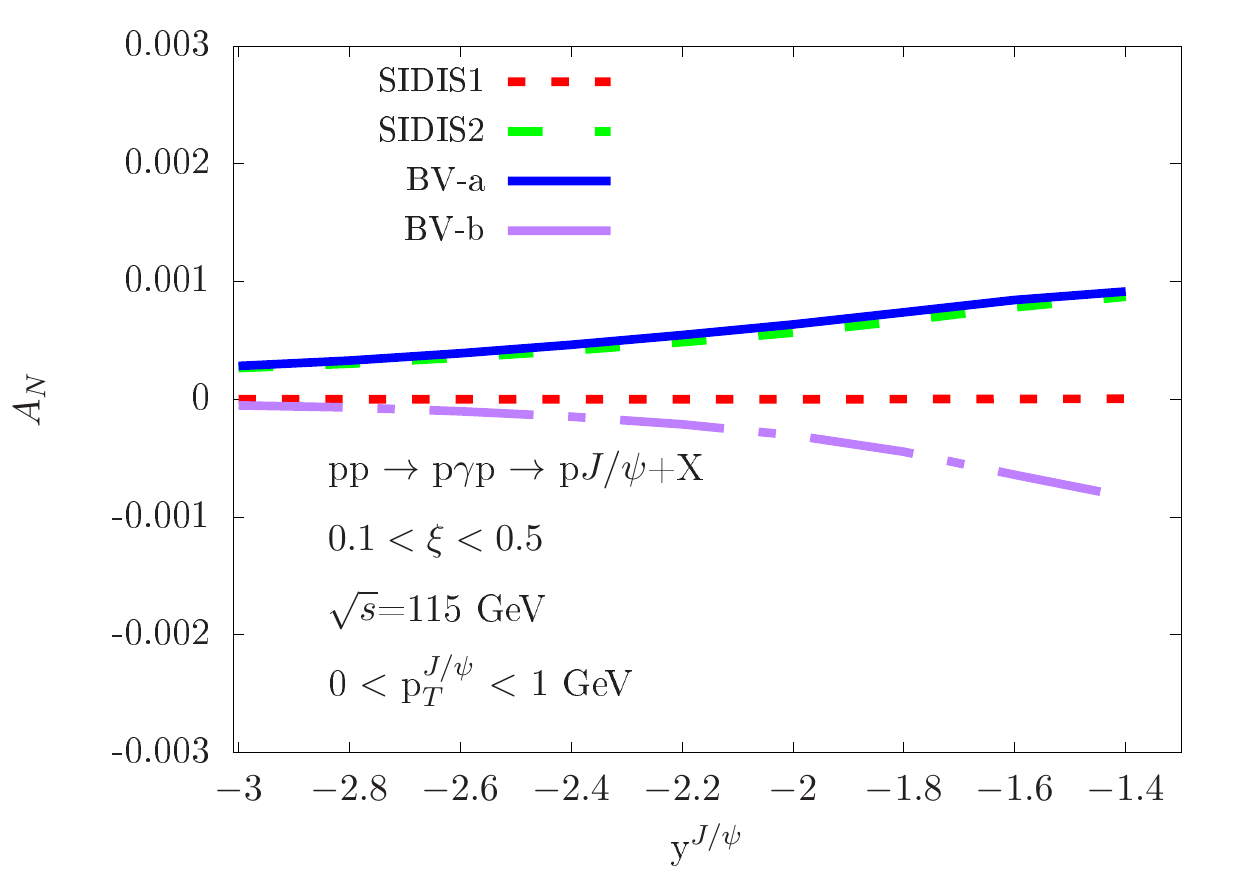}
   \includegraphics[height=4.8cm,width=5.0cm]{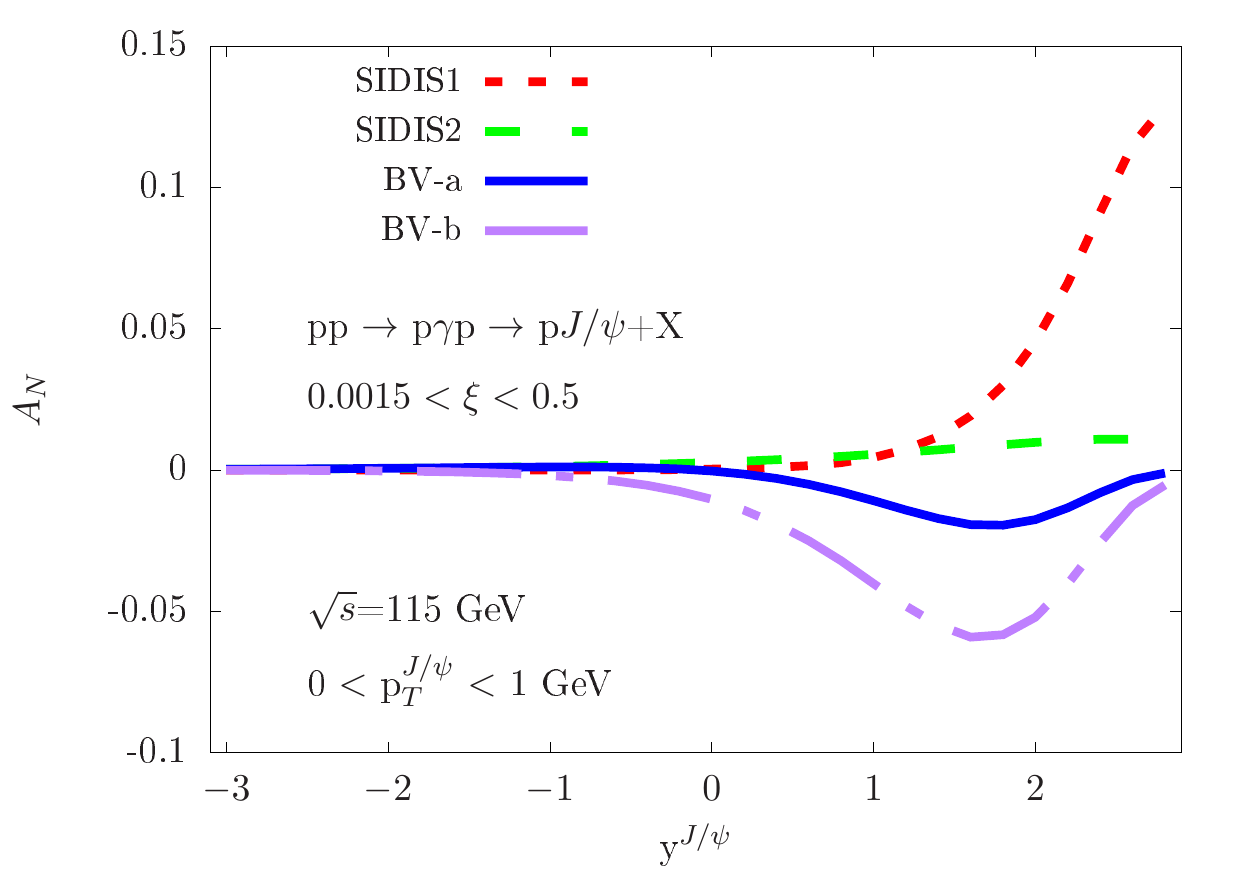}
   \includegraphics[height=4.8cm,width=5.0cm]{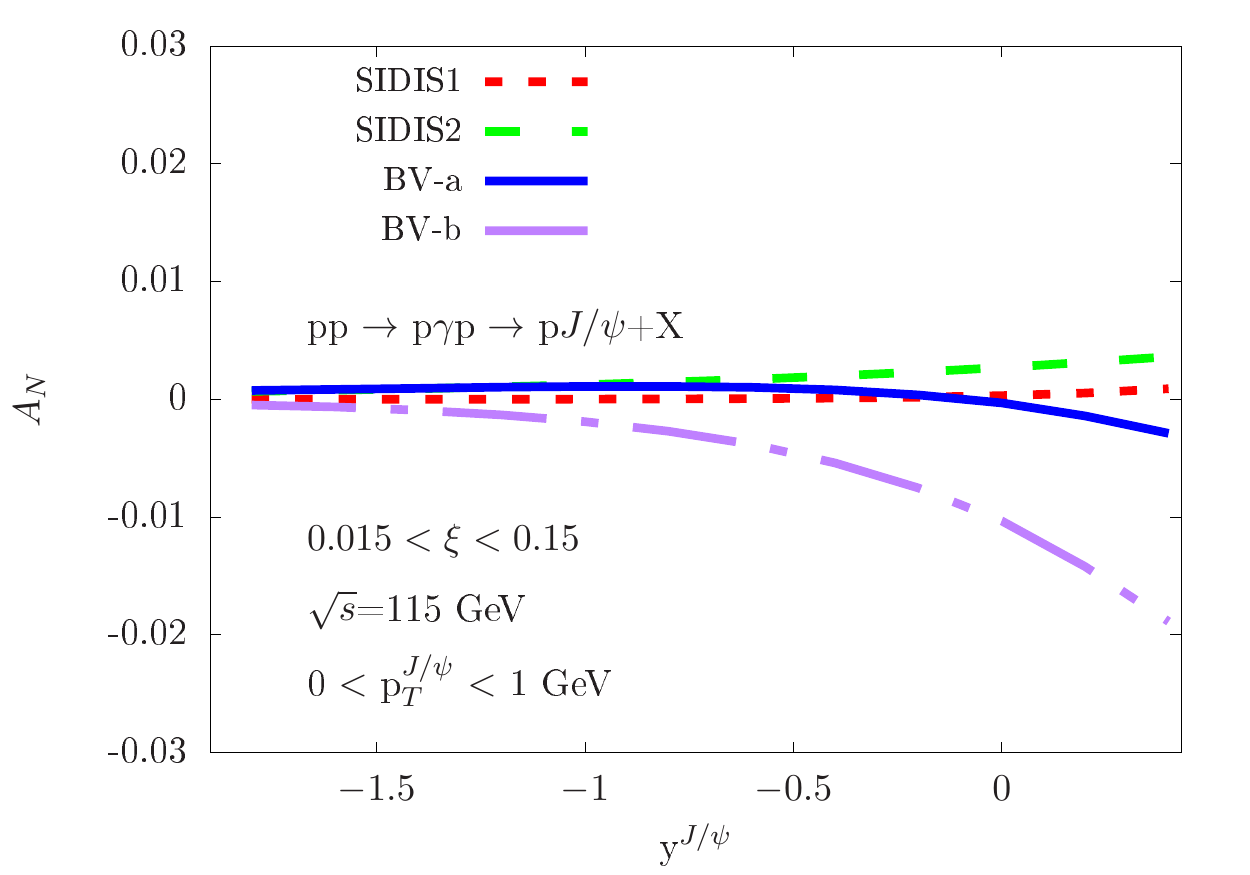}
   \caption{ \normalsize (color online)
Single spin asymmetry in $\rm pp^{\uparrow}\to p\gamma p^{\uparrow}\to p\mathcal{Q}+X$ process
as a function of $\rm y^{J/\psi}$ for $0.1<\xi<0.5$ (left panel), $\rm y^{J/\psi}$ for $0.0015<\xi<0.5$ (middle panel) and $\rm y^{J/\psi}$ for $0.015<\xi< 0.15$ (right panel)
at $\rm \sqrt{s}$ = 115 GeV (AFTER@LHC) using DGLAP (SIDIS1, SIDIS2, BV-a and BV-b).}
\label{fig4:xi}
\end{figure}

\begin{figure}[htp]
\centering
   \includegraphics[height=4.6cm,width=4.4cm]{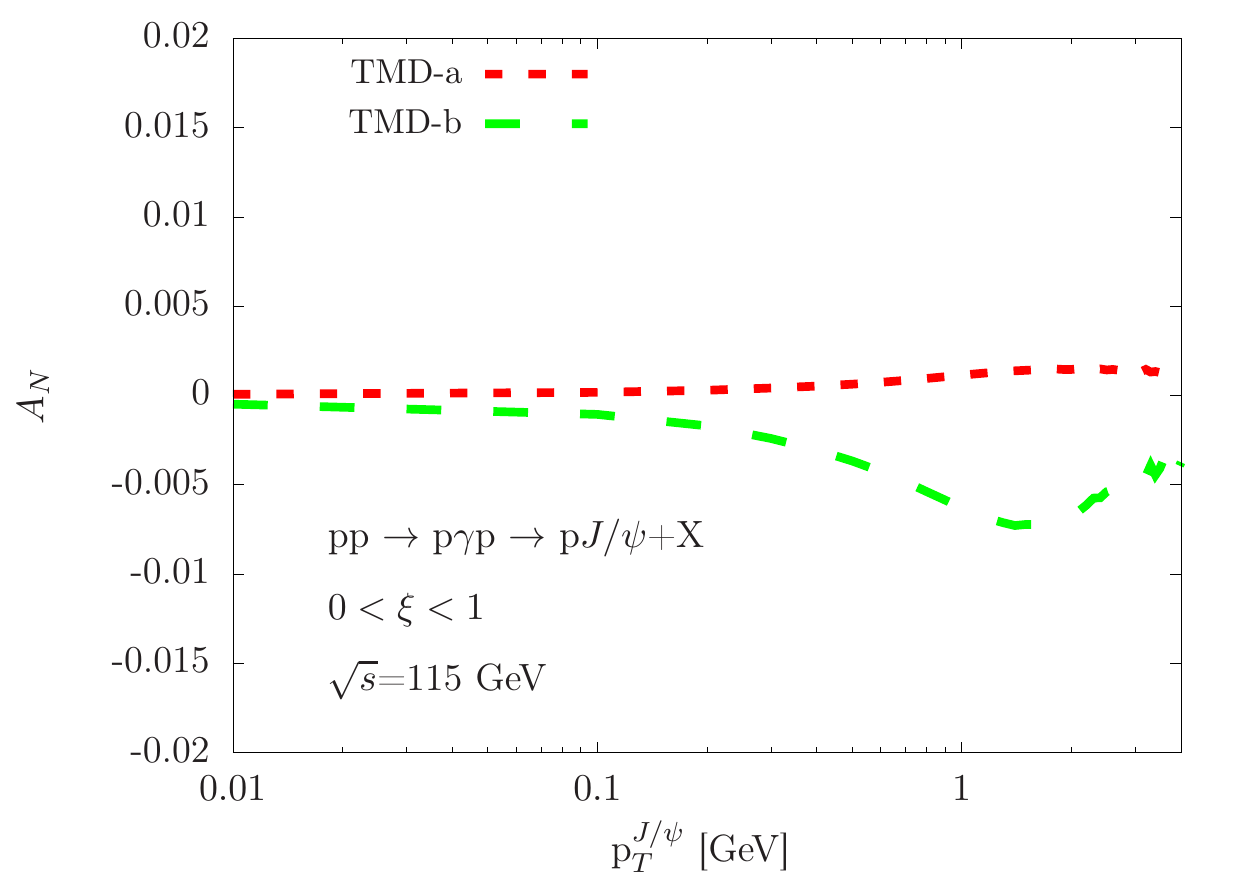}
   \includegraphics[height=4.6cm,width=4.4cm]{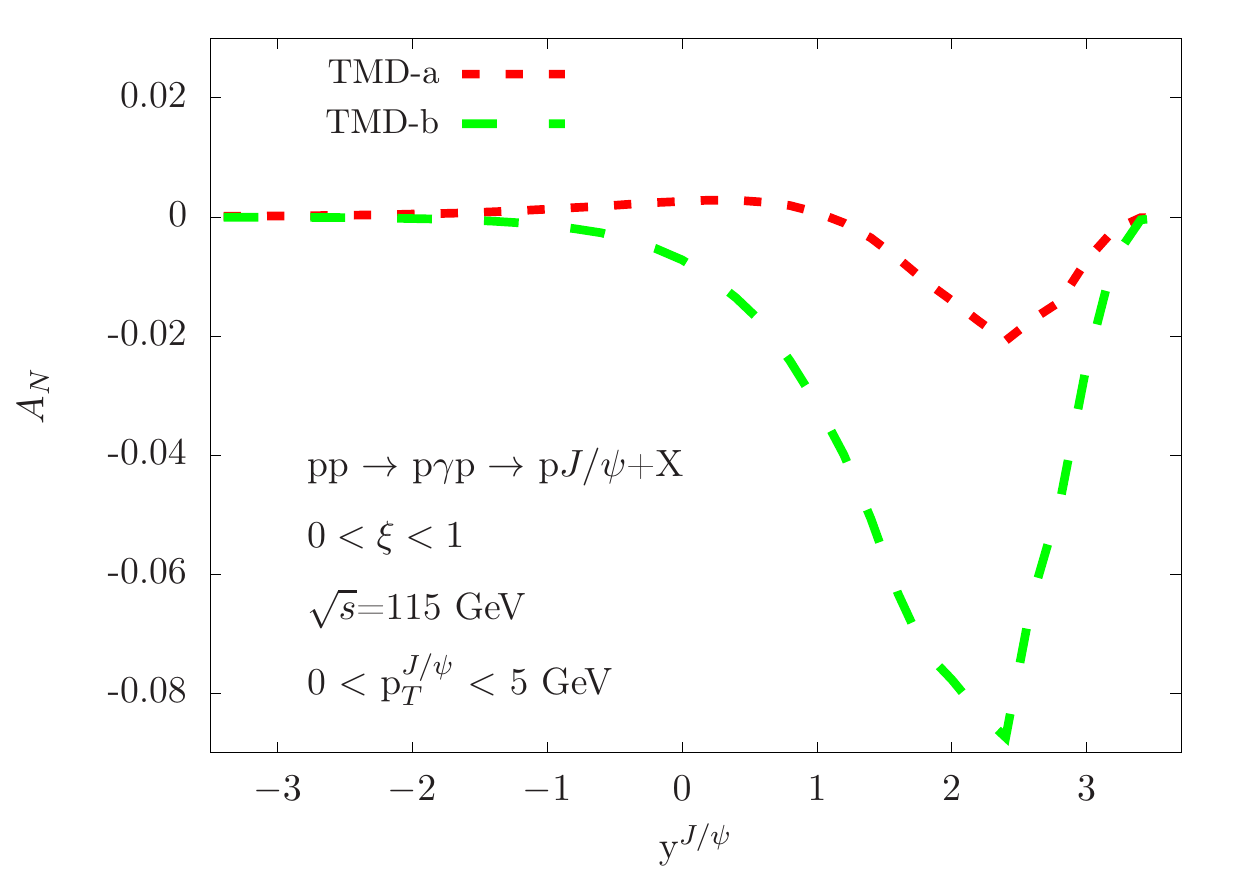}
   \includegraphics[height=4.6cm,width=4.4cm]{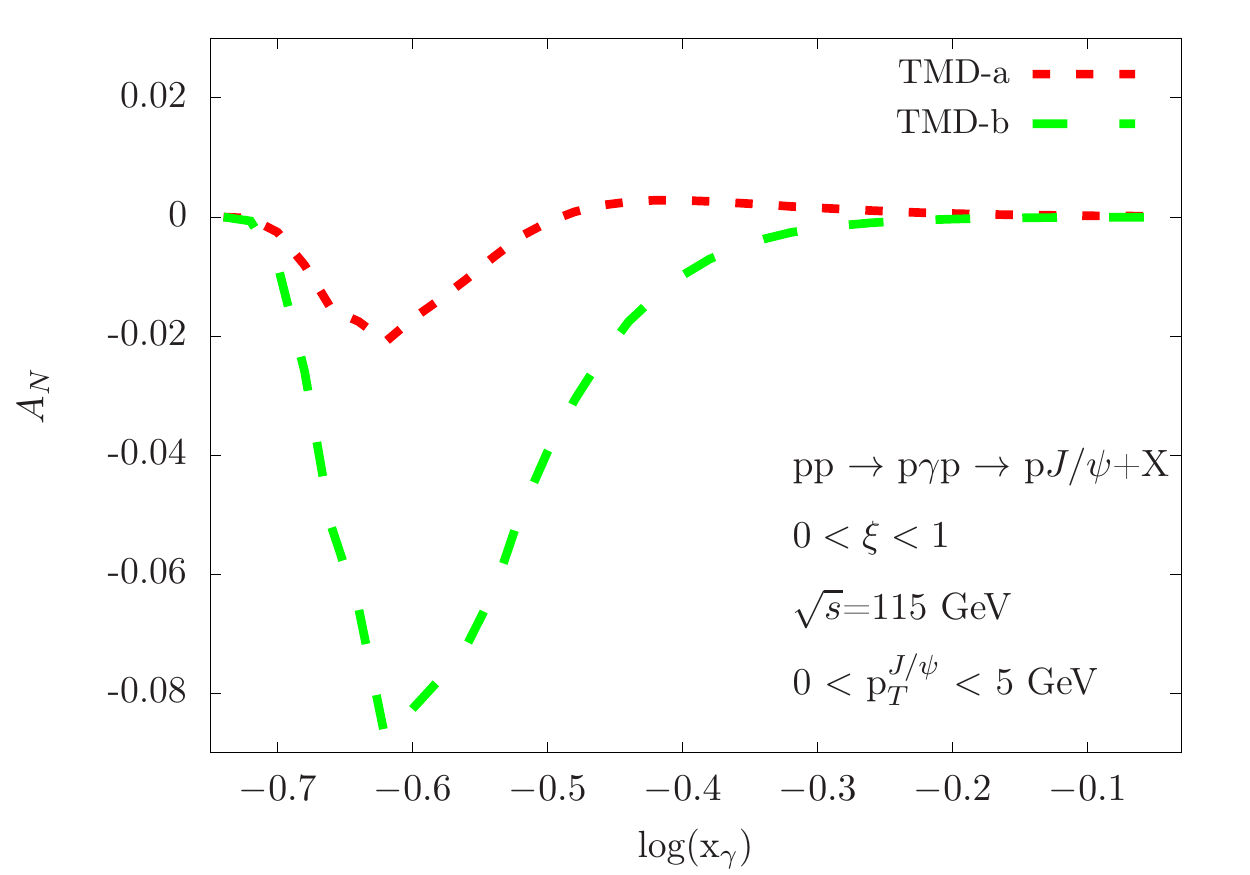}
   \includegraphics[height=4.6cm,width=4.4cm]{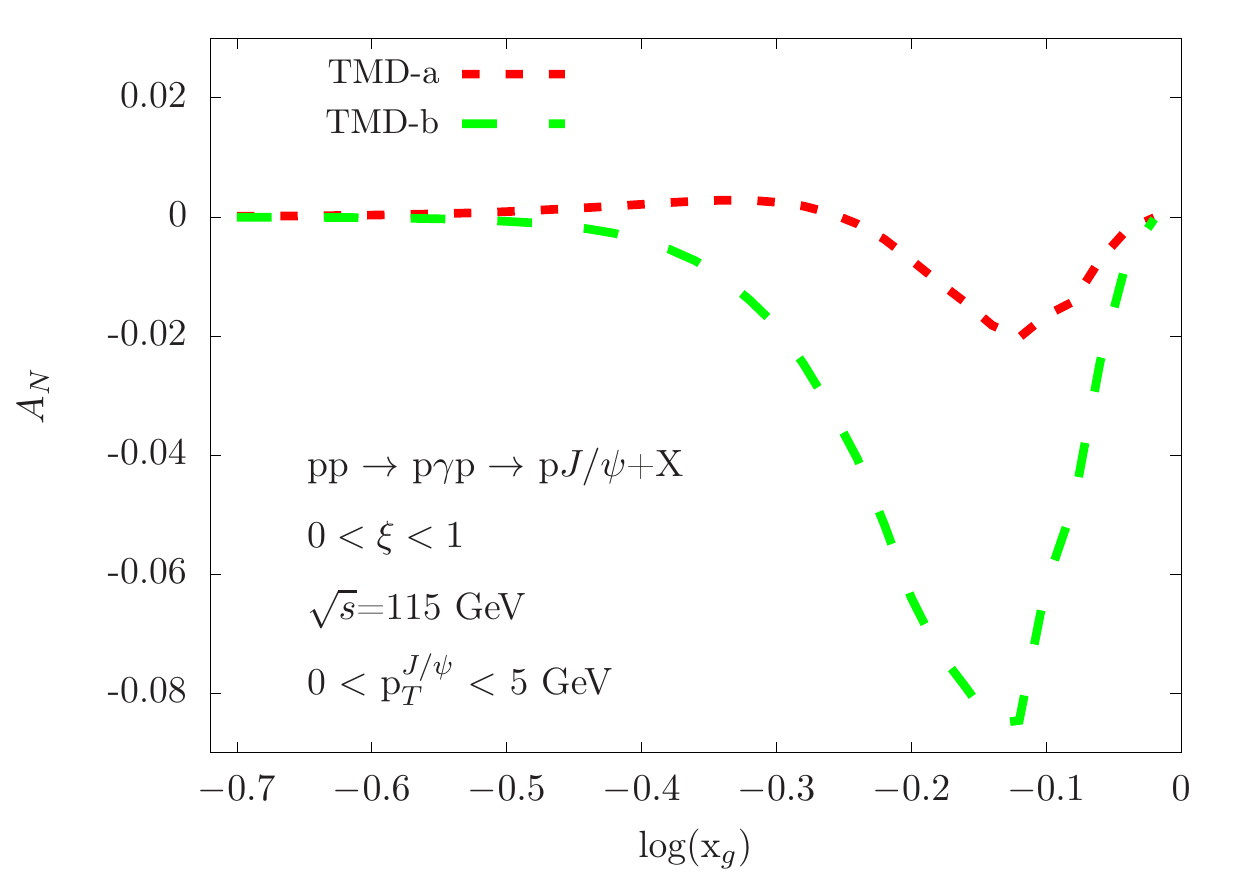}
   \includegraphics[height=4.6cm,width=4.4cm]{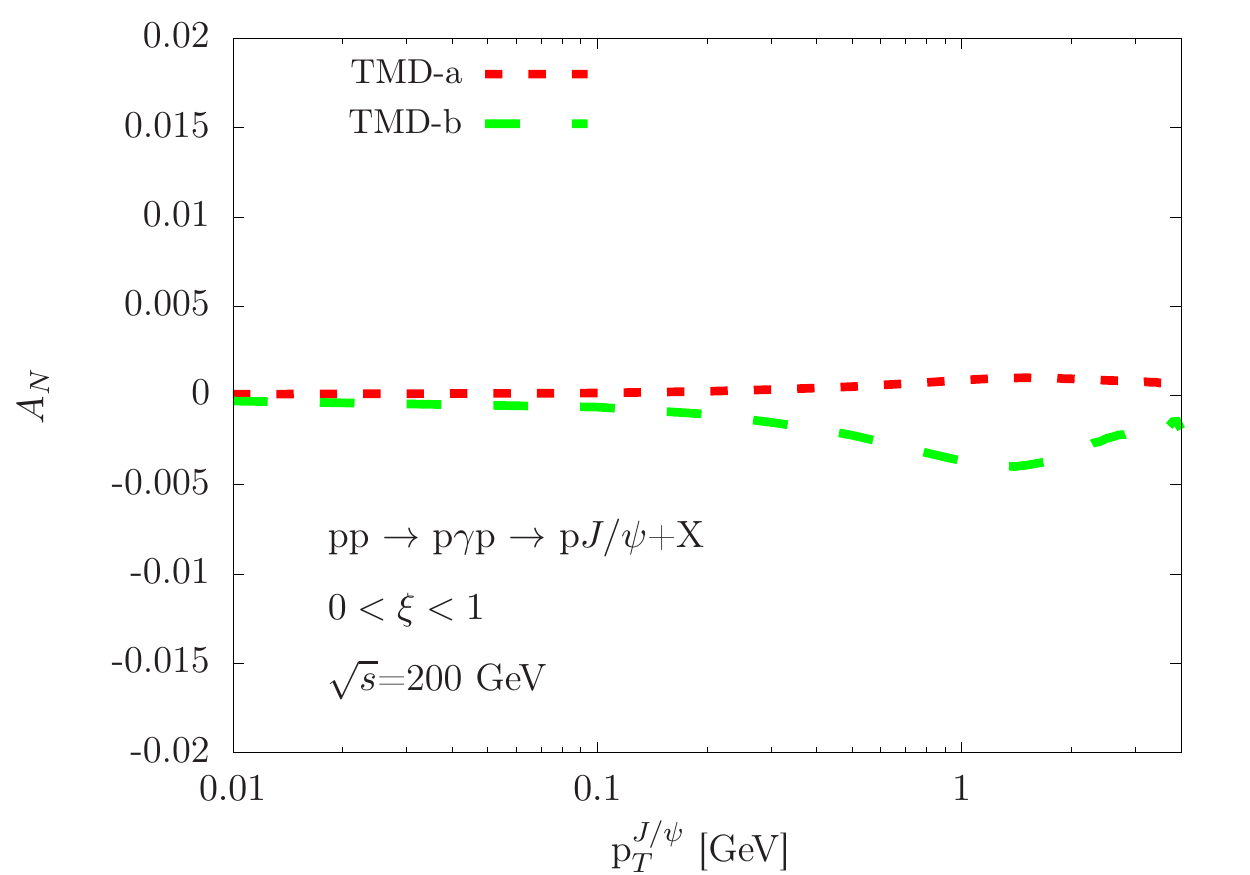}
   \includegraphics[height=4.6cm,width=4.4cm]{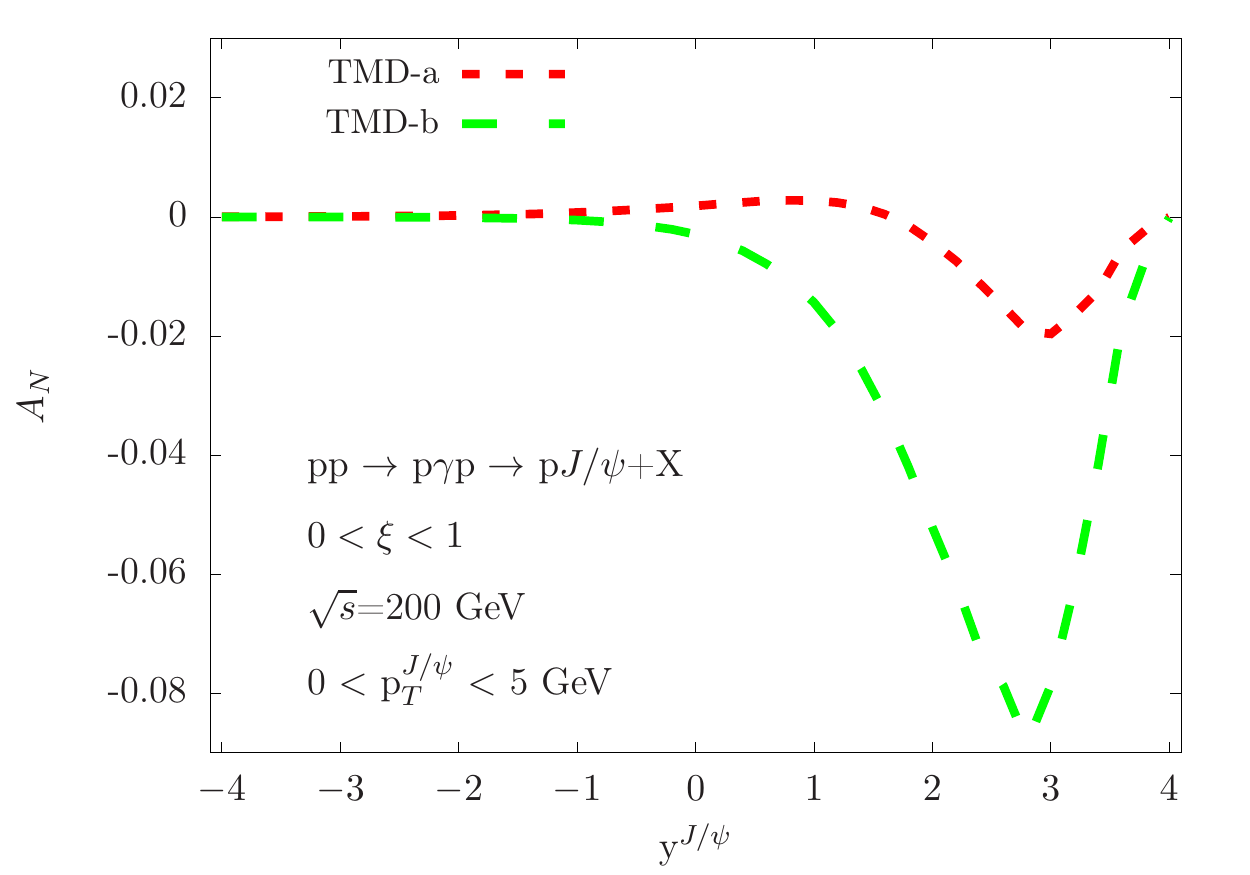}
   \includegraphics[height=4.6cm,width=4.4cm]{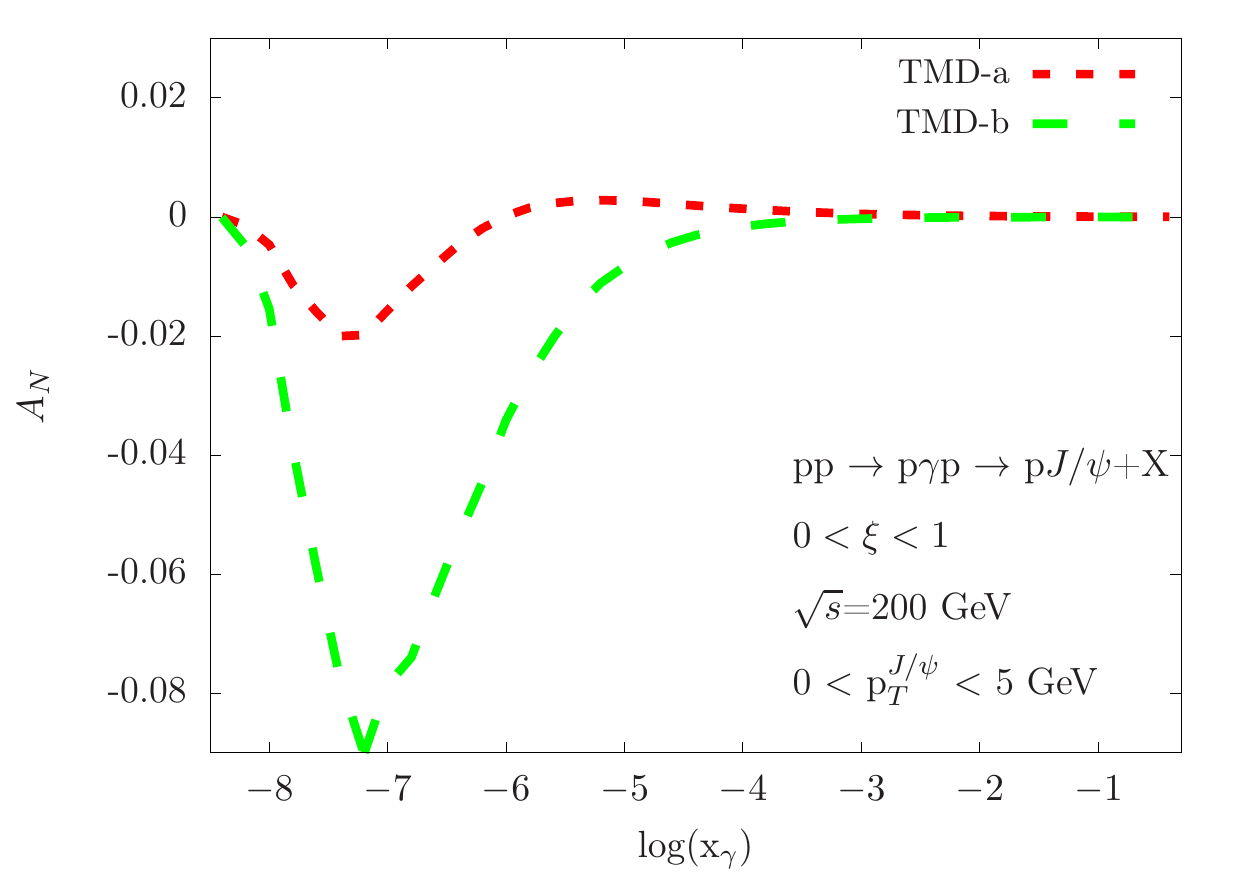}
   \includegraphics[height=4.6cm,width=4.4cm]{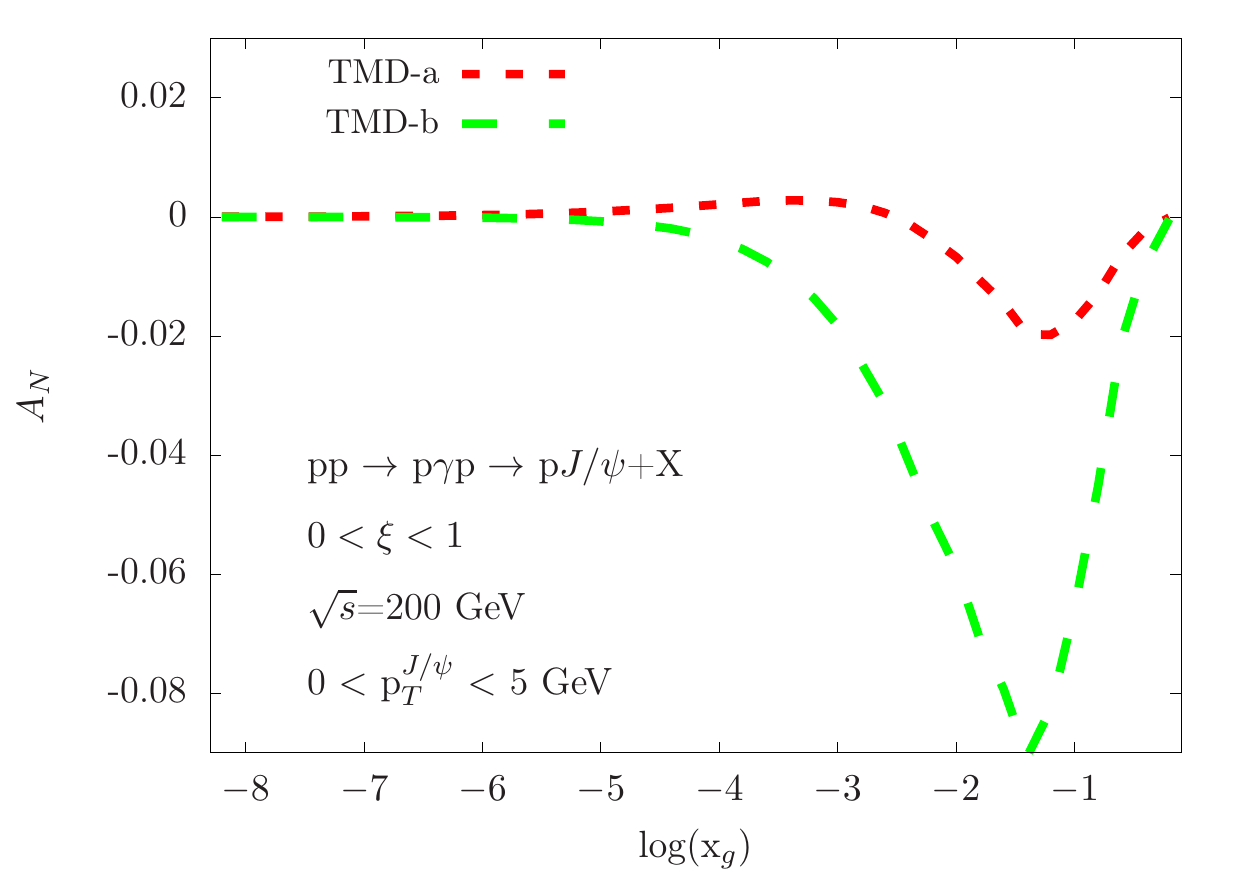}
   \includegraphics[height=4.6cm,width=4.4cm]{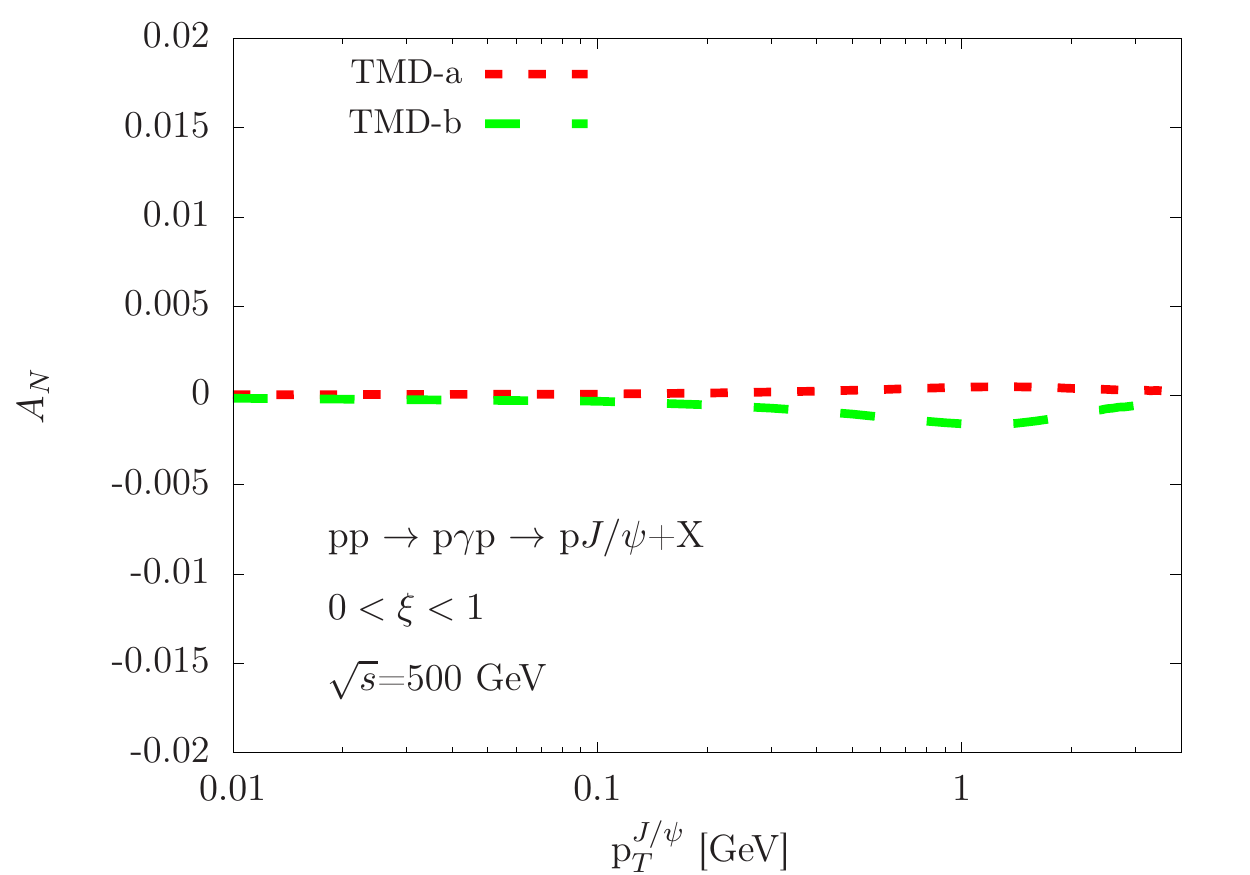}
   \includegraphics[height=4.6cm,width=4.4cm]{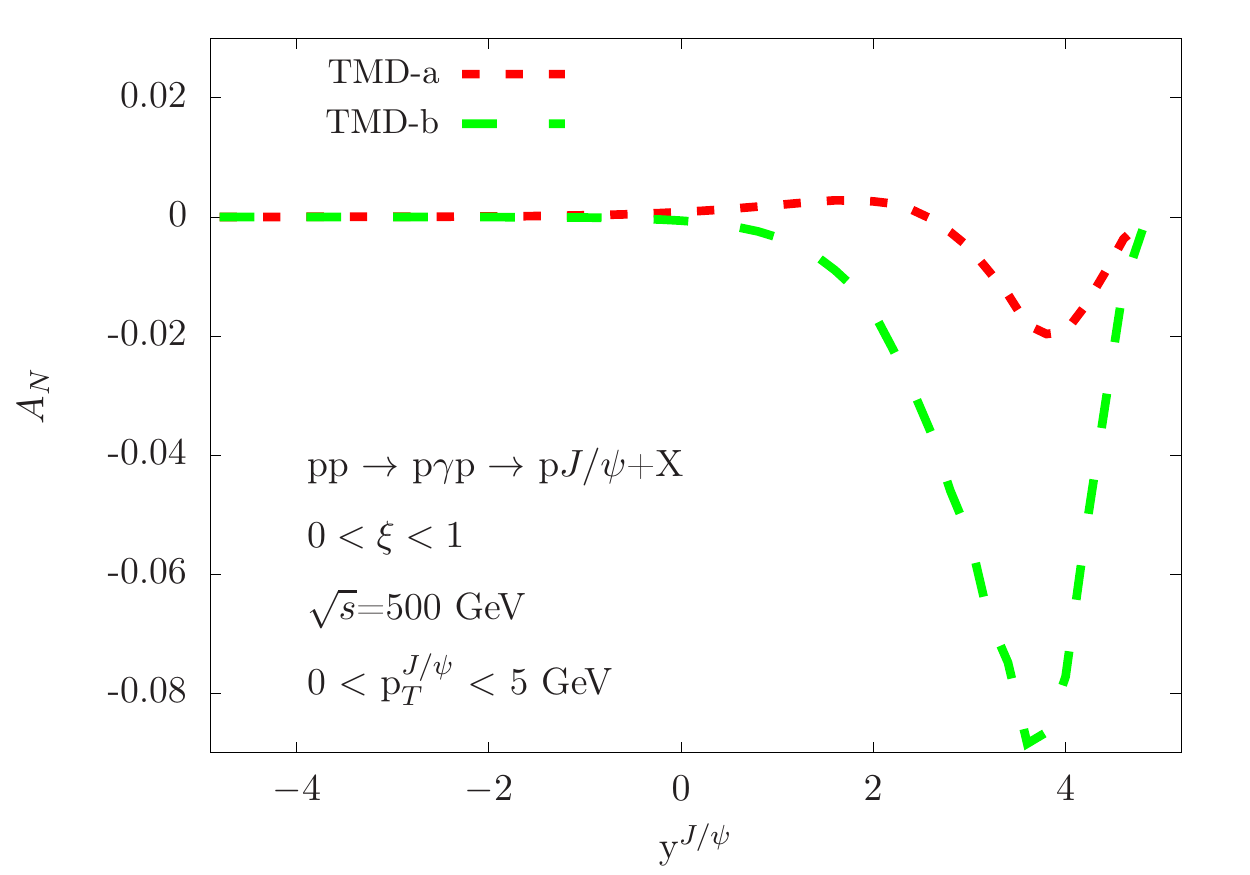}
   \includegraphics[height=4.6cm,width=4.4cm]{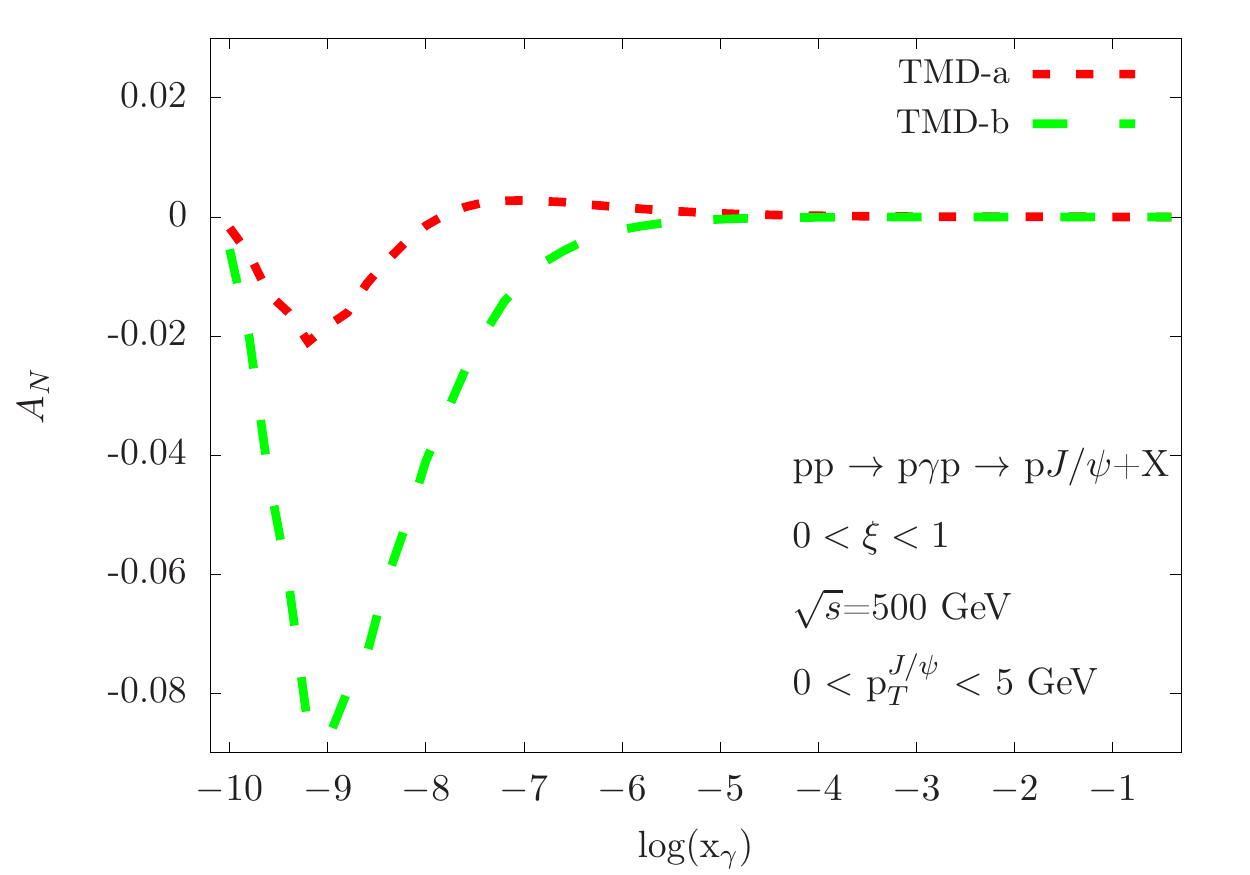}
   \includegraphics[height=4.6cm,width=4.4cm]{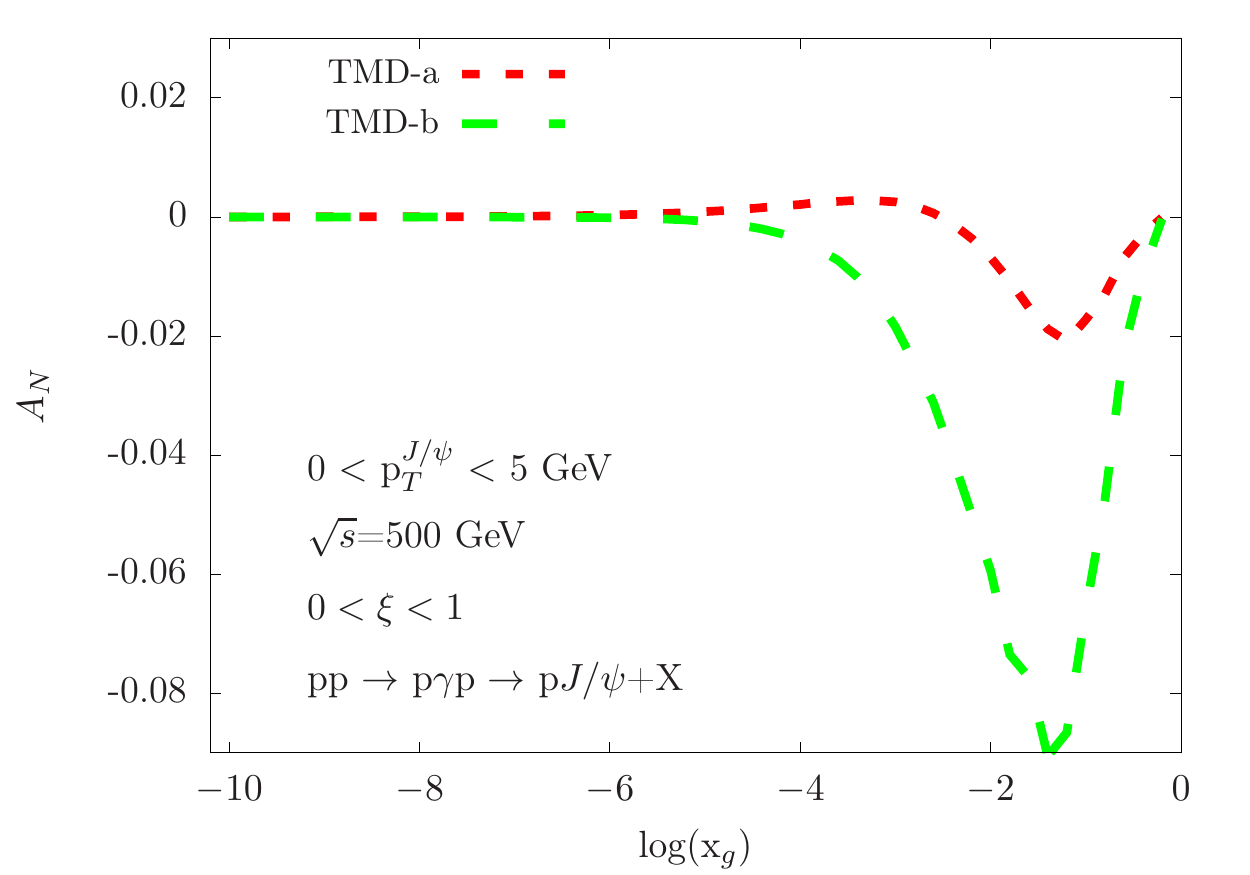}
  \caption{ \normalsize (color online)
Single spin asymmetry in $\rm pp^{\uparrow}\to p\gamma p^{\uparrow}\to p\mathcal{Q}+X$ process as a function of $\rm p_{T}^{J/\psi}$ (left column  panels), $\rm y^{J/\psi}$ (left middle column  panels), $\rm log{(x_{\gamma})}$ (right middle column panels) and $\rm log{(x_{g})}$ (right column panels) at $\rm \sqrt{s}$ = 115 GeV (AFTER@LHC), $\rm \sqrt{s}$ = 200 GeV (RHIC1) and $\rm \sqrt{s}$ = 500 GeV (RHIC2) using TMD (TMD-a and TMD-b).}
\label{fig4:115GeVTMD}
\end{figure}

\begin{figure}[htp]
\centering
   \includegraphics[height=4.8cm,width=5.0cm]{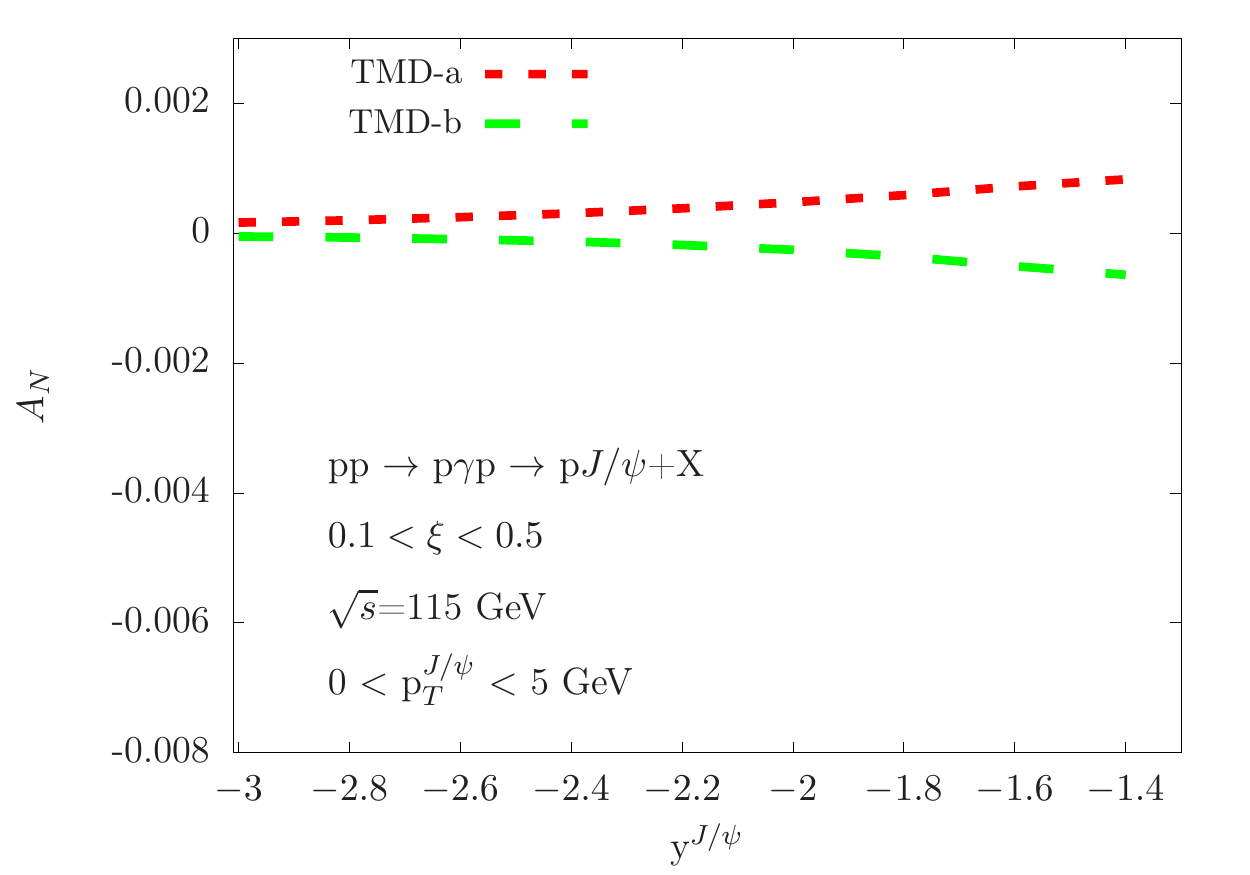}
   \includegraphics[height=4.8cm,width=5.0cm]{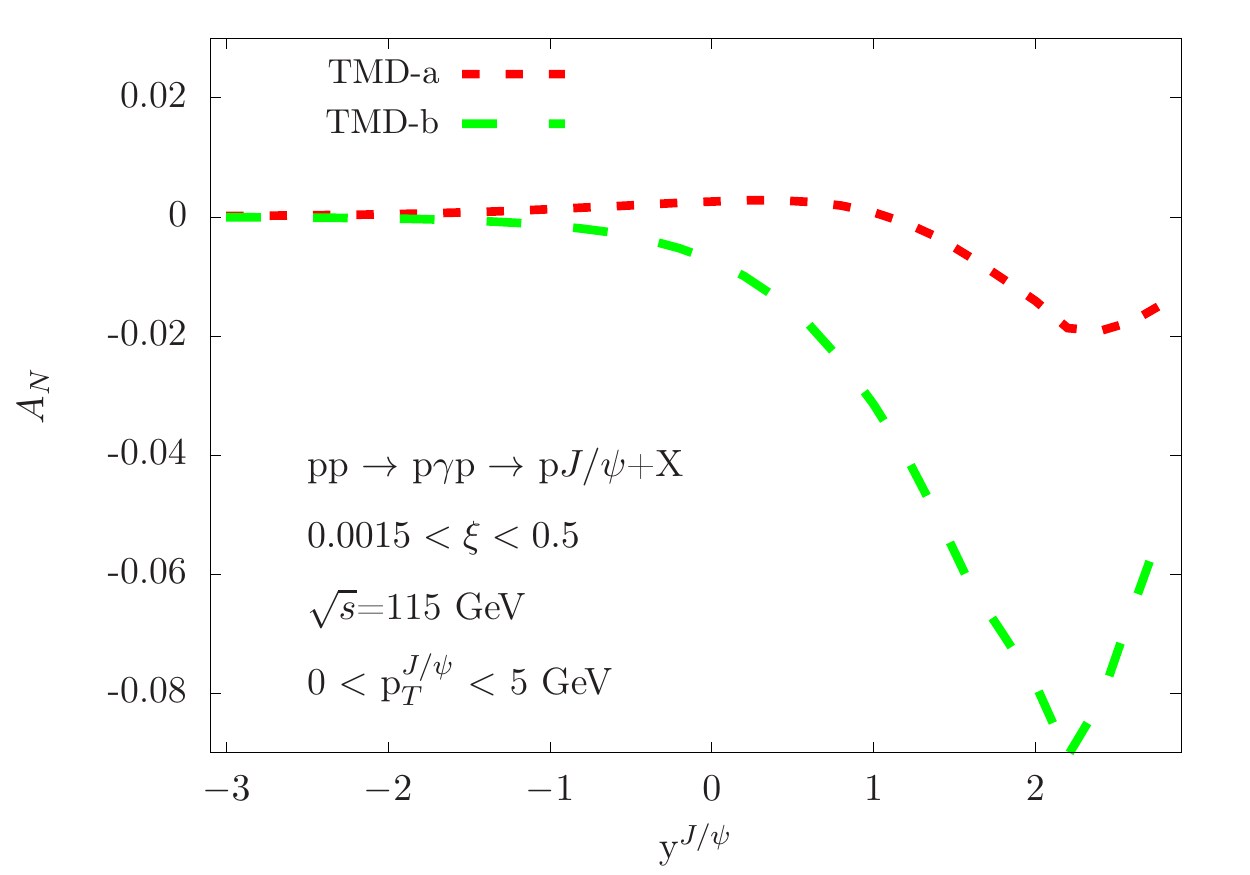}
   \includegraphics[height=4.8cm,width=5.0cm]{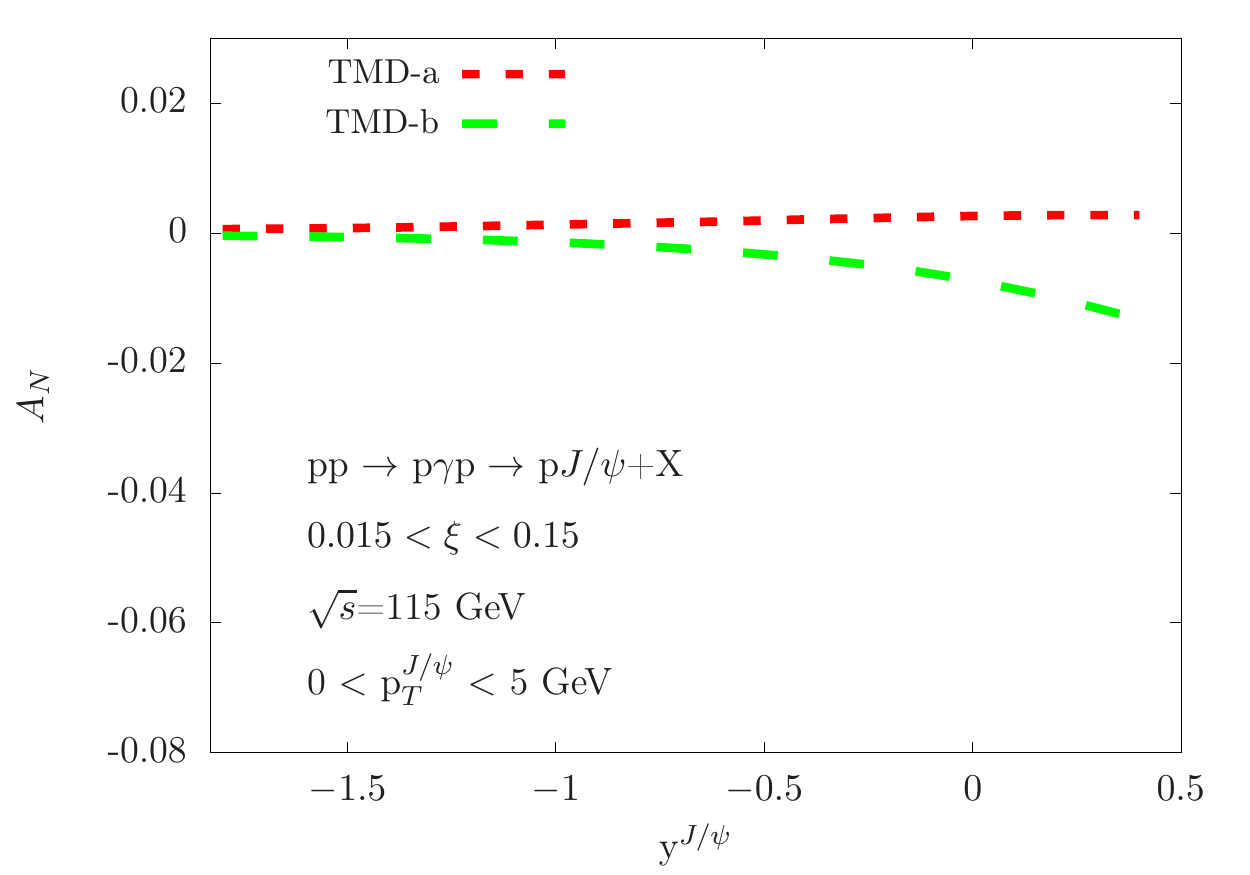}
   \caption{ \normalsize (color online)
Single spin asymmetry in $\rm pp^{\uparrow}\to p\gamma p^{\uparrow}\to p\mathcal{Q}+X$ process as a function of $\rm y^{J/\psi}$ for $0.1<\xi<0.5$ (left panel), $\rm y^{J/\psi}$ for $0.0015<\xi<0.5$ (middle panel) and $\rm y^{J/\psi}$ for $0.015<\xi<0.15$ (right panel) at $\rm \sqrt{s}$ = 115 GeV (AFTER@LHC) using TMD (TMD-a and TMD-b).}
\label{fig4:500GeVTMD}
\end{figure}

\begin{figure}[htp]
\centering
   \includegraphics[height=4.8cm,width=5.0cm]{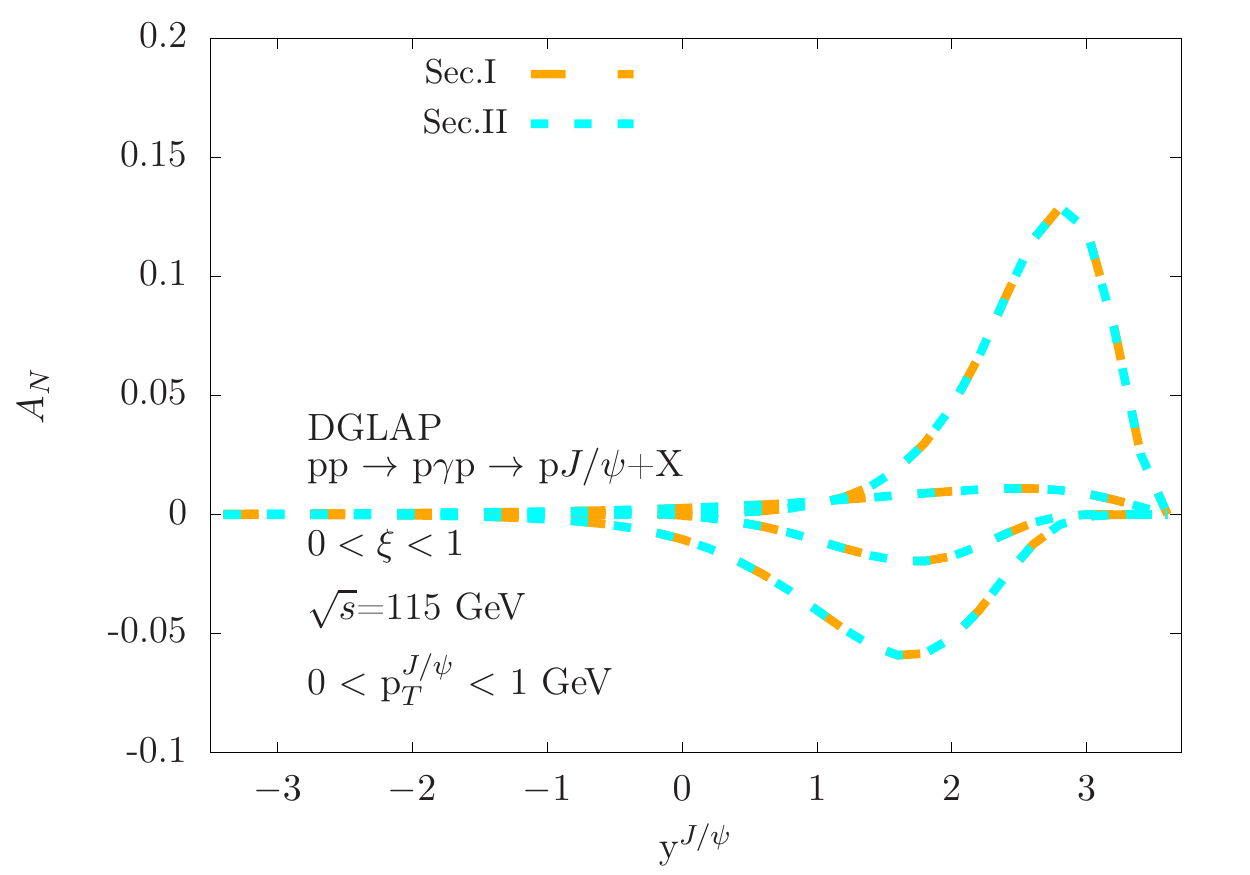}
   \includegraphics[height=4.8cm,width=5.0cm]{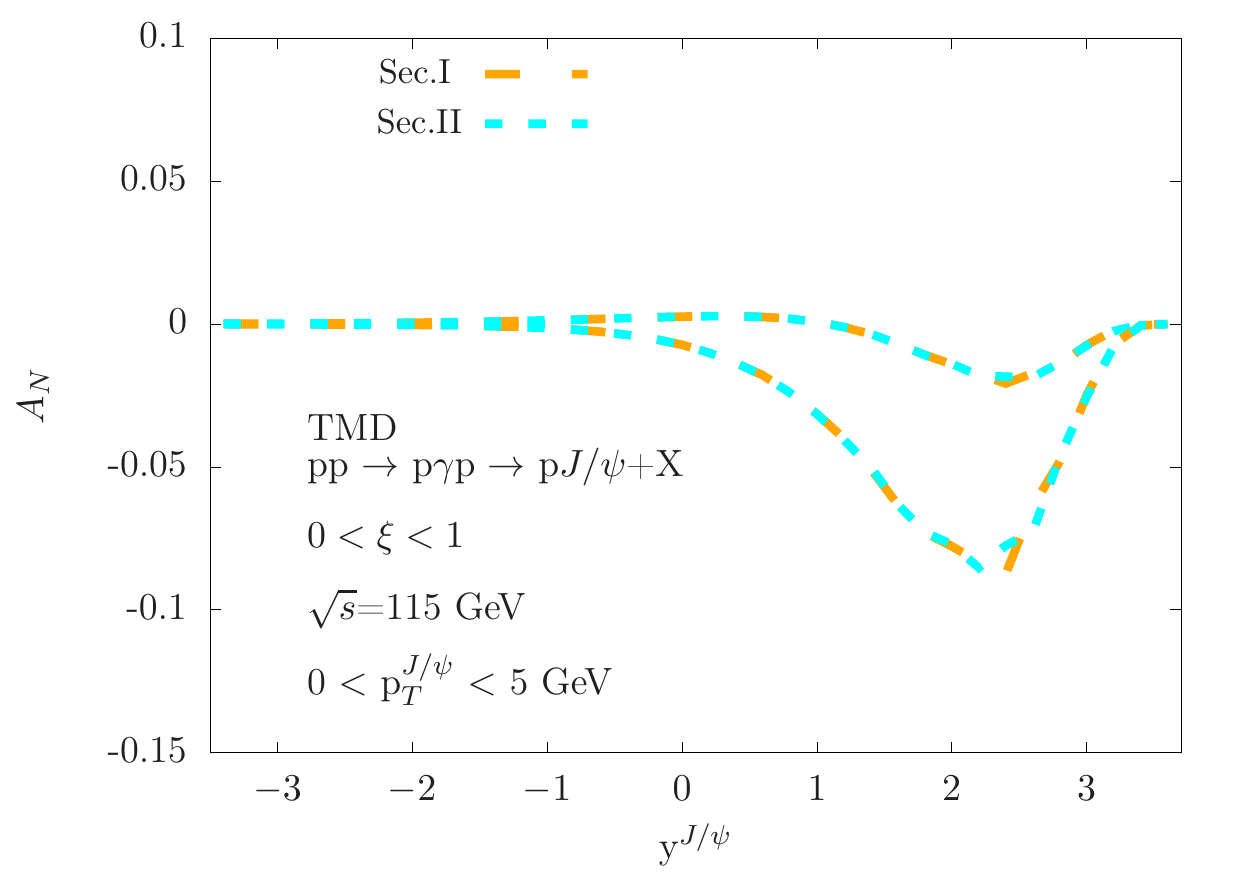}
   \caption{ \normalsize (color online)
 The comparison of single spin asymmetry evaluated by using Set-I and Set-II for DGLAP SSA (left panel) and TMD SSA (right panel) in $\rm pp^{\uparrow}\to p\gamma p^{\uparrow}\to p\mathcal{Q}+X$ process as a function of $\rm y^{J/\psi}$ for $0<\xi<1$ at $\rm \sqrt{s}$ = 115 GeV (AFTER@LHC).}
\label{fig5:115GeVDGLAPandTMDUNC}
\end{figure}

In the following, we investigate the Sivers asymmetries through $\rm J/\psi$ photoproduction in $\rm p^\uparrow p$ collisions with forward proton tagging. At our convenience, the Sivers asymmetry for the different kinematic variables in DGLAP (TMD) evolution are displayed in Fig.\ref{fig4:115GeV}(Fig.\ref{fig4:115GeVTMD}) and as a function of $\rm p_{T}^{J/\psi}$, $\rm y^{J/\psi}$, $\rm log{(x_{\gamma})}$ and $\rm log{(x_{g})}$ respectively,
while in Fig.\ref{fig4:xi}(Fig.\ref{fig4:500GeVTMD}) it is only shown in terms of $\rm y^{J/\psi}$. Furthermore, in Fig.\ref{fig5:115GeVDGLAPandTMDUNC} we analyze the single spin asymmetry with Set-I and Set-II at $\rm \sqrt{s}$ = 115 GeV (AFTER@LHC) in order to get the SSA uncertainty from charmonium productions. The predicted SSAs are sequentially fixed for three distinct center of mass energies $\rm \sqrt{s}$ = 115 GeV (AFTER@LHC), $\rm \sqrt{s}$ = 200 GeV (RHIC1) and $\rm \sqrt{s}$ = 500 GeV (RHIC2) in Figs.\ref{fig4:115GeV} and \ref{fig4:115GeVTMD} whereas the obtained SSAs in Figs.\ref{fig4:xi} and \ref{fig4:500GeVTMD}
are given for the center of mass energy $\rm \sqrt{s}$ = 115 GeV (AFTER@LHC). The configuration of the figures is in this fashion: "SIDIS1" and "SIDIS2" are the representations of the SSA got in DGLAP evolution approach by taking into consideration two sets of best fit parameters SIDIS1 and SIDIS2 using D'Alesio at el. \cite{DAlesio:2015fwo} fit parameters of GSF; the "BV-a" and "BV-b" plots are obtained by employing Anselmino et al. \cite{Anselmino:2016uie} fit parameters.

From the hard process calculation of $\rm c\overline{c}$ pair production through 2 $\to$ 1 partonic process, it has been noticed that the Fock states are only produced in color octet, that is to say, the asymmetry by means of $\rm J/\psi$ formation being very receptive to the production machinery will be non zero in color octet contribution and zero in singlet contribution \cite{Yuan:2008vn}. The $\rm d\Delta\sigma$ involving the polarized cross sections and $\rm 2d\sigma$ including the unpolarized ones of Eq.\ref{AN} are computated when the initial heavy quark pair is produced in color octet state. Despite of different shapes of curves, distinct kinematic variables in DGLAP evolution and being in the factoriazation validity in the range of $\rm x_{g}$, the SSA is declining as the center of mass energy of the experiment is rising. In the all range of the forward detector acceptance, $0<\xi<1$, as shown in Figs.\ref{fig4:115GeV} and \ref{fig4:xi}, the SSA versus $\rm p_{T}^{J/\psi}$, $\rm y^{J/\psi }$, $\rm log{(x_{\gamma})}$ and $\rm log{(x_{g})}$ do have two realms of opposite signs, positive and negative, as estimated by SIDIS and BV parameters. The obtained asymmetry as function of $\rm p_{T}^{J/\psi}$, $\rm y^{J/\psi}$, $\rm log{(x_{\gamma})}$ and $\rm log{(x_{g})}$ using "SIDIS1" and "SIDIS2" parameters are positive whereas those of "BV-a" and "BV-b" parameters are negative. The sign of the asymmetry depends on relative magnitude of $\rm N_u$ and $\rm N_d$ and these have opposite sign which can be observed in TABLE.\ref{table1}. The magnitude of $\rm N_d (x_g)$ is dominant compared to $\rm N_u (x_g)$
as a result the asymmetry is negative. Nevertheless, the magnitude and sign of the asymmetry strongly depends on the modeling of GSF.

As shown in Fig.\ref{fig4:115GeV}, the obtained asymmetry as function of $\rm p_{T}^{J/\psi}$ using "BV-a" parameters is near zero although the center of mass energy is unequal, while the obtained asymmetry as function of $\rm y^{J/\psi }$, $\rm log{(x_{\gamma})}$ and $\rm log{(x_{g})}$ using "SIDIS2" parameters is close to zero despite the fact that the center of mass energy is also different. The assessed asymmetry using "BV-b" is maximal around 5\% as function of $\rm p_{T}^{J/\psi }$ at $\rm \sqrt{s}$ = 115 GeV (AFTER@LHC). It has also notably remarked that the estimated asymmetry by utilizing "SIDIS1" is leading around 12.5\% as function of $\rm y^{J/\psi }$, $\rm log{(x_{\gamma})}$ and $\rm log{(x_{g})}$ for the three different experiments suggested at LHC forward detector acceptance. Our attention has been purposefully drawn by the fact that the SSA peak value has been displaced positively and negatively at the right along the $\rm y^{J/\psi }$ and $\rm log(x_{g})$-axes respectively, with the rise of $\rm \sqrt{s}$ when the SSA is presented as a function of rapidity and $\rm log(x_{g})$. As an explanation, there exists a dependence between gluon momentum fraction and the rapidity given by this formula: $\rm x_{g}=\frac{Me^{+y}}{\sqrt{s}}$ where $\rm M$ is the mass of $\rm J/\psi$. The proportionality coefficient, $\rm x_g^{\alpha} (1-x_g)^{\beta}$ or Sivers effect, gives the ratio of SSA against rapidity. Even though the same behavior has been observed in the plot of SSA versus $\rm log{(x_{\gamma})}$, the left displacement of SSA peak value is negative along the axis with increase of $\rm \sqrt{s}$. The reason is that there is also a linear correlation between the photon momentum fraction and the forward detector acceptance $\xi$, and indirectly with SSA. The SSA peak displacement value of $\rm log{(x_{\gamma})}$ and $\rm log{(x_{g})}$ distributions are on the left and right respectively, but they remain negative. In DGLAP evolution, the $\rm y^{J/\psi }$, $\rm log{(x_{\gamma})}$ and $\rm log{(x_{g})}$ distributions are more sensitive to measurement of SSA than that of $\rm p_{T}^{J/\psi }$ one which tends to zero. We comment here that the Gaussian ansatz being $\rm k_{\bot}$-dependence and factorized from $\rm x$-dependence is not suitable to study the SSA \cite{Su:2014wpa} in low-$\rm x_{g}$ region, and needs to be modified to survive \cite{Nadolsky:1999kb,Nadolsky:2000ky} as we have mentioned above.

In TMD evolution at the LHC forward detector acceptance $0<\xi<1$, as seen in Fig.\ref{fig4:115GeVTMD}, the asymmetry with respect to $\rm p_{T}^{J/\psi }$ using "TMD-a" is zero and positive whilst that of "TMD-b" parameters is also zero and negative. At curved lines, the asymmetry slightly and positively (negatively) escape from zero using "TMD-a" ("TMD-b"). Their effects are diametrically apposite. As for the $\rm y^{J/\psi }$, $\rm log{(x_{\gamma})}$ and $\rm log{(x_{g})}$ distributions, asymmetries are negative and slightly run away from zero. Asymmetry with regard to $\rm p_{T}^{J/\psi}$ obtained from "TMD-a" and "TMD-b" set parameters are more at AFTER@LHC experiment ($\rm \sqrt{s}$ = 115 GeV).
SSA peak value displacement in TMD evolution is almost similar to that in DGLAP evolution
for the representation of assessed asymmetries versus $\rm y^{J/\psi}$, $\rm log{(x_{\gamma})}$ and $\rm log{(x_{g})}$, and asymmetry signs are also the same for "TMD-a" and "TMD-b" parametrization corresponding to "BV-a" and "BV-b" parametrization. The predicted SSA peak value in DGLAP evolution is around 12.5\% compare to that of TMD evolution which is around 7.8\%.


In Fig.\ref{fig4:xi}, we have noticed that the behaviors of SSA versus $\rm y^{J/\psi}$ are utterly different from the right panel to the left panel due to the forward detector acceptance range and DGLAP parametrizations. The obtained asymmetry as function of $\rm y^{J/\psi}$ using "BV-b" parameters is negative, and positive using "SIDIS2" and  "SIDIS1" parameters for all three forward detector acceptances. The strangeness by employing "BV-a" parameters comes from the sign change of SSA and it is positive for the left panel and, negative for the middle and right panels. For the left panel and right panel, the obtained asymmetries using "SIDIS1" are zero. As for the  right panel, the obtained asymmetry using "BV-a" is zero and for the middle panel, the assessed asymmetry employing "SIDIS2" is also zero. The assessed asymmetry using "SIDIS1" is maximal around 12.5\% as function $\rm y^{J/\psi}$ for the middle panel.

In Fig.\ref{fig4:500GeVTMD}, the forward detector acceptance range and TMD parametrizations do also influence the evaluated asymmetries. The curves for the right  and left panels exhibit almost the same behavior whereas the curve for the middle panel shows the maximal value of the assessed asymmetry using "TMD-b" parameters around 7.8\%, and asymmetries are negative using TMD parameters. The asymmetries for the right and left panels using "TMD-a" parameters are positive and lightly run from zero while for "TMD-b" are negative and lightly run from zero, too. The left and the right panels of Figs.\ref{fig4:xi} and \ref{fig4:500GeVTMD} have their SSA peak values occurring at small rapidities, and the maximum and minimum of their SSA peak values are smaller than of the middle panels arising at large rapidities. The shape, the sign and the value of SSA in both evolutions are dissimilar because of their parametrizations.

We have estimated the SSAs from two different LDMEs denoted by Set-I and Set-II in DGLAP evolution as well as TMD evolution in Fig.\ref{fig5:115GeVDGLAPandTMDUNC}, at $\rm \sqrt{s}$ = 115 GeV (AFTER@LHC) considering all range of the detector acceptance as an example. We find that the differences between the SSA for Set-I and Set-I are small and this means that analyzed uncertainties are also pretty small between the two sets. Based on our numerical estimation, we have found that the uncertainties are of order of $10^{-3}$ thus negligible, this is understood by the fact that the SSAs are calculated through the ratio of the polarized cross sections to the unpolarized ones of $\rm J/\psi$ photoproduction in its expression, therefore the uncertainties arising from the charmonium production are independent on the LDMEs or even the PDFs. The order of uncertainty also remains small at $\rm \sqrt{s}$ = 200 GeV (RHIC1) and $\rm \sqrt{s}$ = 500 GeV (RHIC2) as they are almost independent on colliding energies.

\section{SUMMARY AND DISCUSSIONS}
\label{summary}

In this paper, we have evaluated the magnitude of single spin asymmetry in photoproduction of $\rm J/\psi$ by resorting to NRQCD approach, considering both DGLAP evolution and TMD evolution. Sizable asymmetry is predicted as a function of $\rm y^{J/\psi }$, $\rm log{(x_{\gamma})}$ and $\rm log{(x_{g})}$ respectively. The maximal value of single spin asymmetry is about 12.5\% for DGLAP evolution and 7.8\% for TMD evolution. The minimum and maximun of SSA are almost independent of energy. The obtained asymmetry as a function of $\rm y^{J/\psi}$ and $\rm log{(x_{g})}$, and the obtained asymmetry as a function of $\rm log{(x_{\gamma})}$ show opposite displacement of their peaks. We choose three different forward detector acceptances, and find that $0.0015<\xi< 0.5$ is the region where most of the SSA effects are kept and possiblely detected for both DGLAP and TMD evolutions with our choice of parametrization. In summary, our results point out that the magnitude of the asymmetry can be estimated by photoproduction of $\rm J/\psi$ with forward detector acceptances at the RHIC and AFTER@LHC experiments. 

\begin{acknowledgments}
Hao Sun is supported by the National Natural Science Foundation of China (Grant No.11675033) and by the Fundamental Research Funds for the Central Universities (Grant No. DUT18LK27).
\end{acknowledgments}

\bibliography{v1}

\end{document}